\documentclass[12pt]{article}
\usepackage{graphicx}
\usepackage{pdfpages}
\usepackage{adjustbox}
\usepackage{amsmath,amsfonts,amssymb,amsthm}
\usepackage{caption}
\usepackage[ruled,vlined]{algorithm2e}
\usepackage{bbm}
\usepackage{breakcites}
\usepackage{tabularx}
\usepackage{geometry}
\usepackage{enumitem}
\usepackage{natbib}
\usepackage{authblk}
\usepackage{graphicx}
\usepackage{subcaption}
\usepackage{hyperref}
\usepackage{ulem}

\setlist[itemize]{align=parleft,left=0pt..1em}
\newcolumntype{L}{>{\raggedright\arraybackslash}X}

\usepackage{stix}
\usepackage{color}

\theoremstyle{definition}

\newtheorem{theorem}{Theorem}[section]

\newtheorem{lemma}[theorem]{Lemma}

\newcommand{\independent}{\perp \!\!\! \perp}

\newcommand{\bY}{\mathbf{Y}}
\newcommand{\bV}{\mathbf{V}}

\newcommand{\bX}{\mathbf{X}}
\newcommand{\bZ}{\mathbf{Z}}
\newcommand{\bF}{\mathbf{F}}

\newcommand{\bx}{\mathbf{x}}
\newcommand{\bz}{\mathbf{z}}
\newcommand{\by}{\mathbf{y}}
\newcommand{\bv}{\mathbf{v}}

    \title{Using Machine Learning to Test Causal \\ Hypotheses in Conjoint Analysis\thanks{We thank Naoki Egami for advice. Imai thanks the Alfred P. Sloan Foundation for partial support (Grant \# 2020--13946). Ham and Janson were partially supported by a CAREER grant from the National Science Foundation (Grant \# DMS-2045981).} }
    
    \author[1]{Dae Woong Ham\thanks{Email: \href{mailto:daewoongham@g.harvard.edu}{\tt daewoongham@g.harvard.edu}}} 
    \author[1,2]{Kosuke Imai\thanks{Email: \href{mailto:imai@harvard.edu}{\tt imai@harvard.edu}, URL: \url{https://imai.fas.harvard.edu}}}
    \author[1]{Lucas Janson\thanks{Email: \href{mailto:ljanson@fas.harvard.edu}{\tt ljanson@fas.harvard.edu}, URL: \url{http://lucasjanson.fas.harvard.edu}}}
    \affil[1]{\textit{Department of Statistics, Harvard University}}
    \affil[2]{\textit{Department of Government, Harvard University}}
    
    \date{\today}
    
\begin{document}
    \maketitle

\begin{abstract}
   Conjoint analysis is a popular experimental design used to measure multidimensional preferences. Researchers examine how varying a factor of interest, while controlling for other relevant factors, influences decision-making. Currently, there exist two methodological approaches to analyzing data from a conjoint experiment. The first focuses on estimating the average marginal effects of each factor while averaging over the other factors.  Although this allows for straightforward design-based estimation, the results critically depend on the distribution of other factors and how interaction effects are aggregated. An alternative model-based approach can compute various quantities of interest, but requires researchers to correctly specify the model, a challenging task for conjoint analysis with many factors and possible interactions.  In addition, a commonly used logistic regression has poor statistical properties even with a moderate number of factors when incorporating interactions.  We propose a new hypothesis testing approach based on the conditional randomization test to answer the most fundamental question of conjoint analysis: Does a factor of interest matter {\it in any way} given the other factors? Our methodology is solely based on the randomization of factors, and hence is free from assumptions.  Yet, it allows researchers to use any test statistic, including those based on complex machine learning algorithms.  As a result, we are able to combine the strengths of the existing design-based and model-based approaches.  We illustrate the proposed methodology through conjoint analysis of immigration preferences and political candidate evaluation. We also extend the proposed approach to test for regularity assumptions commonly used in conjoint analysis. An open-source software package is available for implementing the proposed methodology.\footnote{The proposed methodology is implemented via an open-source software R package \textit{CRTConjoint}, available through the Comprehensive R Archive Network (\url{https://cran.r-project.org/package=CRTConjoint}).}
\end{abstract}

\newpage
\section{Introduction}

Conjoint analysis, introduced more than half a century ago \citep{conjoint}, is a factorial survey-based experiment designed to measure preferences on a multidimensional scale. Under a commonly used ``forced-choice'' design, respondents are presented with two alternative profiles of randomly selected attributes (e.g., candidates with different traits or products with different features).  They are then asked to choose their preferred profile.  Conjoint analysis has been extensively used by marketing firms to determine desirable product characteristics \citep[e.g.,][]{market, market2}.  Recently, it has gained popularity among social scientists \citep{AMCE, socialscience2} who are interested in studying individual preferences concerning elections \citep[e.g.,][]{gender}, immigration \citep[e.g.,][]{immigration}, employment \citep[e.g.,][]{employment}, and other issues.

When analyzing conjoint experiments, the {\it design-based} approach, pioneered by \citet{AMCE}, has been by far the most popular among social scientists.  The main advantage of this nonparametric approach is its simplicity---it uses the difference-in-means estimator or linear regression to infer the average marginal component effect (AMCE) of each factor by averaging over the distribution of other factors. However, because the AMCE makes inferences about the marginal effects
averaged over all the other factors, it may fail to capture important interactions. This is potentially problematic given that practitioners tend to use AMCE-based confidence intervals that are narrow and contain zero to conclude that a factor has a weak causal effect \citep{AMCE, immigration, gender}. However, a narrow AMCE-based confidence interval containing zero only implies that a factor has a weak \textit{marginal} effect, not necessarily that its total causal influence is weak.

A possible solution is the {\it model-based} approach that ranges from traditional parametric regression models \citep{mcfa:73, marketgeneral,marketreg} to more recent machine learning algorithms \citep{egam:imai:19,distributional_effects,dispute2,dispute3,gopl:imai:pash:22}.  While this approach can efficiently estimate various quantities of interest, it has the potential drawback of model misspecification, producing biased inference. Although researchers can reduce model misspecification by adopting a more complicated model (e.g., adding interaction terms of increasing order), such an approach can substantially reduce statistical power and produce invalid $p$-values even when the model is correctly specified \citep{phasetransition}.  While subgroup analysis, a common practice to analyze only a subset of the data, is simpler, it suffers from well-known problems of multiple testing and $p$-hacking, which are of serious concern in conjoint analysis given a large number of possible causal effects of interest \citep[see, e.g.,][]{p_hacking1, p_hacking2}.  Finally, the use of machine learning algorithms, which is becoming increasingly common, cannot yield even consistent estimates in high-dimensional settings without strong assumptions.

In this paper, we propose a new approach to analyzing data from conjoint analysis that combines the strengths of the existing design-based and model-based approaches (Section~\ref{section:methodology}).  Specifically, we show how to conduct assumption-free hypothesis testing based on the conditional randomization test \citep[CRT;][]{CRT}.  In the causal inference literature, the CRT has been used to test interference between units \citep{aron:12,athe:eckl:imbe:18} and in other causal applications such as genetic studies \citep{CRT_genetic}. Instead of focusing on a particular causal effect, we ask the most fundamental question of conjoint analysis: Does a factor of interest matter {\it in any way} given the other factors? The proposed approach allows one to answer this question with greater statistical power than the AMCE by utilizing flexible machine learning algorithms but without making \emph{any} assumption about the underlying causal structure (e.g., presence or absence of interaction effects, patterns of heterogeneous effects, or within-respondent correlation across profile comparisons).  Despite its flexibility, the CRT has an attractive statistical property that the resulting $p$-values are \emph{exactly} valid regardless of the sample size or the number of factors. We also extend the proposed approach to test whether specific factor levels of interest, rather than all levels of a factor, influence respondent preferences \textit{in any way}  (Section~\ref{subsection:generalization}).

The proposed methodology can also test the validity of assumptions commonly invoked in conjoint analysis \citep{AMCE} (Section~\ref{subsection:extensions}). First, we show how to test for the presence of the profile order effect.  Under the forced conjoint design, for example, reversing the order in which two profiles are presented within each evaluation may change the choice of profile. Second, we test the assumption of no carryover effect, which states that each respondent's response only depends on the current profiles and is not affected by previous evaluations. This assumption may be violated if respondents learn over several profile evaluations. Third, we test the assumption of no fatigue effect, which precludes the possibility that as respondents evaluate more profiles, they get tired and answer questions differently \citep{choice_task, satisficing}. Thus, the proposed hypothesis testing approach can serve as the first step of analyzing conjoint data without assumptions, complementing existing approaches that estimate causal quantities of interest. 

After presenting the proposed methodology, we conduct simulation studies in Section~\ref{section:simulations} to show that the CRT can achieve a higher statistical power than the AMCE by exploiting machine learning algorithms to detect complex treatment interactions. For empirical illustration, we apply the proposed methodology to two existing conjoint analyses (Section~\ref{section:empirical_results}). The first pertains to immigration preferences among United States (U.S.) citizens. While some researchers contend that U.S. citizens generally prefer high-skilled immigrants regardless of their countries of origin, others have suggested that racially prejudiced respondents discriminate against non-European immigrants \citep{immigration,immigration_debate}.  By combining machine learning algorithms with the CRT, we find that respondents do differentiate according to whether immigrants are from Mexico or European countries. The second application considers the role of gender in candidate evaluation \citep{gender}.  The original analysis found that voters discriminate between male and female candidates only for presidential elections but not for Congressional elections.  However, a recent study suggests that this finding about congressional candidates may critically depend on the distribution of other characteristics such as partisanship and policy positions \citep{distributional_effects}. We apply the proposed methodology and show that gender does play a statistically significant role in voter evaluation of Congressional candidates. 

We emphasize that the CRT complements other existing methods by offering a useful test of whether a factor of interest matters at all.  The test can be useful even when the AMCE fail to detect any statistically significant result.

\section{Empirical Application}
\label{section:empirical_examples}

In this section, we briefly describe two empirical applications concerning the role of ethnocentrism in immigration preferences and gender discrimination in political candidate evaluations.  We also outline the limitations of the commonly used approach based on the AMCE to motivate the proposed methodology.  In Section~\ref{section:empirical_results}, we revisit these applications and apply our hypothesis testing approach. 

\subsection{Role of Country of Origin in Immigration Preference}
\label{subsection:immigration}

Immigration is one of the most contentious issues in the United States today.  A large body of literature investigates how cultural, economic, and racial factors shape public attitudes towards immigration \citep[see, e.g.,][and references therein]{hain:hopk:14}.  In an influential study, \citet{immigration} use a conjoint analysis to empirically examine the immigrant characteristics favored or disfavored by U.S. citizens. The survey was fielded between December 2011 and January 2012 on a nationally representative sample of U.S. citizens. The study used the forced-choice design, in which each respondent was presented with a pair of hypothetical immigrant profiles and asked which immigrant they would ``personally prefer to see admitted to the United States.''  Each of 1,396 respondents rated 5 pairs of profiles. 

An immigrant profile consists of nine factors---prior trips to the U.S., reason for application, country of origin, language skills, profession, job experience, employment plans, education level, and gender, each of which has multiple levels (see Table~\ref{tab:immigration_factorlevel} of Appendix~\ref{appendix:data_table} as well as the original article for details).  Most factors are independently and uniformly randomized across their levels with the exception of two restrictions to avoid implausible pairs. First, immigrant profiles that list \textit{escape persecution} as the ``reason of immigration'' can only have \textit{Iraq}, \textit{Sudan}, or \textit{Somalia} as their ``country of origin''.  Second, a high-skill ``profession'' such as \textit{financial analyst}, \textit{research scientist}, \textit{doctor}, and \textit{computer programmer} is possible only if the ``education level'' is at least \textit{2 years of college}. This restricted randomization scheme induces dependencies between these factors that must be properly accounted for when analyzing the data. The survey also contains information about respondents' age, education, ethnicity, gender, and ethnocentrism. The study contains a random sample of 14,018 profiles.

\begin{figure}[t!]
\begin{center}
\includegraphics[width=8cm]{"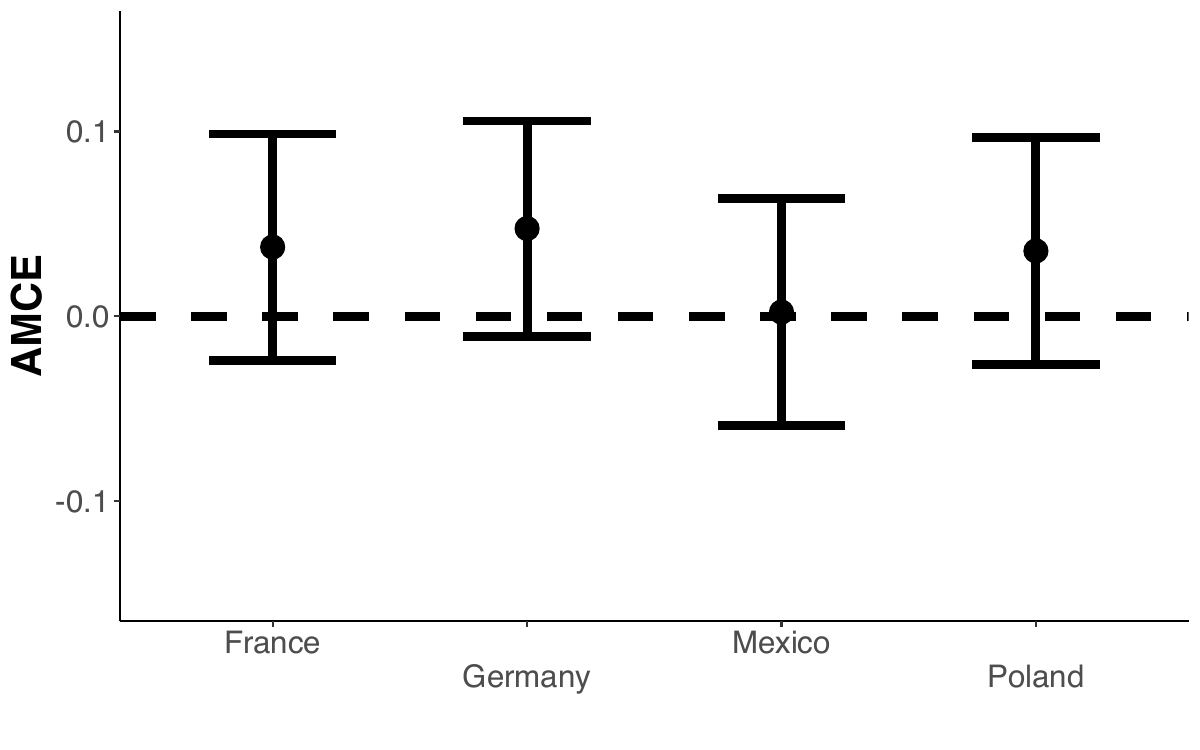"}
\caption{The estimated Average Marginal Component Effects (AMCEs) of immigrants' countries of origin in the \cite{immigration} study. The plot shows the estimated AMCEs for {\it France}, {\it Germany}, {\it Mexico}, and {\it Poland}, which represent the average differences in the estimated probability of choosing an immigrant profile with a specific level of the ``country of origin'' factor, marginalizing other attributes.  The baseline factor level is {\it India}, and the 95\% confidence intervals are also shown.}
\label{fig:immigration_prior_results}
\end{center}
\end{figure}

In this study, \citeauthor{immigration} estimate the AMCE, which represents the marginal effect of a factor of interest averaging over the other factors. Based on the statistically insignificant estimates for the AMCEs of the ``country of origin'' factor for Mexico and European countries (reproduced in Figure~\ref{fig:immigration_prior_results}), they conclude that ``despite media frames focusing on low-skilled, unauthorized immigration from Mexico, there is little evidence of penalty specific to Mexicans.'' (p. 539). The authors obtain these estimates by fitting a linear regression model, where the outcome variable indicates whether the profile is selected and the predictors are the nine randomized factors. To account for the restricted randomization, they also include two sets of interaction terms, one between ``country of origin'' and ``reason of immigration'' and the other between ``profession'' and ``education level''. To obtain the estimated AMCE of \textit{Germany}, for example, \citeauthor{immigration} take the main effect of \textit{Germany} (the baseline is \textit{India}) and then add it to the average of all the interaction terms between \textit{Germany} and the ``reason of immigration'' factor. Clustered standard errors are computed by clustering on each respondent to account for dependency within a respondent.  

Despite this overall finding, the AMCE-based approach may mask relevant interactions and heterogeneous treatment effects.  Indeed, \citeauthor{immigration} conduct a subgroup analysis and find that the ``country of origin'' factor has statistically significant interactions with the respondents' ethnocentrism.  They measure ethnocentrism using the feeling thermometer score (ranging from $0$ to $100$) for the respondent's own groups minus the average feeling thermometer across the other groups.  In addition, \cite{immigration_debate} reanalyze the same dataset and estimate three-way interactions among respondents' ethnocentrism, ``country of origin'', and ``profession''. The authors compute the AMCEs of high-skilled immigrants (baseline of \textit{janitor}) separately for each country of origin and respondent's ethnocentric group.  They find that these AMCEs are different between Mexican and European immigrants when compared among highly ethnocentric respondents \citep[see Figure 1 in][for further details]{immigration_debate}.

In this paper, we apply the proposed hypothesis testing approach to testing whether or not immigrants from Mexico and those from Europe are viewed differently in {\it any way} while controlling for all the other experimental factors as well as the respondent characteristics. The rejection of this null hypothesis would mean that the country of origin of an individual plays a statistically significant role in some United States citizens' preferences about that individual's immigration to the United States.

\subsection{Role of Gender in Political Candidate Evaluation}
\label{subsection:gender}

Perhaps the most common political science application of conjoint analysis is the measurement of voters' candidate preferences.  Recently, several scholars have used conjoint analysis to study the role of gender discrimination in candidate evaluation \citep[e.g.,][]{gender,teel:kall:rose:18}.  We revisit the study by \citet{gender} which examines whether voters prefer candidates of one gender over those of another after controlling for other candidate characteristics.\footnote{This study treats gender as a binary factor with levels {\it Male} and {\it Female}.} The study is based on a sample of voting-eligible adults in the U.S. collected in March 2016 and also uses the forced-choice conjoint design. The following 13 factors are independently and uniformly randomized across their levels: gender, age, race, family, experience in public office, salient personal characteristics, party affiliation, policy area of expertise, position on national security, position on immigrants, position on abortion, position on government deficit, and favorability among the public (see Table~\ref{tab:gender_factorlevel} in Appendix~\ref{appendix:data_table} and the original article for details). The survey also contains information about the respondents' educational background, gender, age, region, social class, partisanship, political interest, and ethnocentrism. There were 1,583 respondents each given 10 tasks, resulting in 15,830 observations, half of which were for congressional candidates and the other half for presidential candidates.

\begin{figure}[t!]
\begin{center}
\includegraphics[width=8cm]{"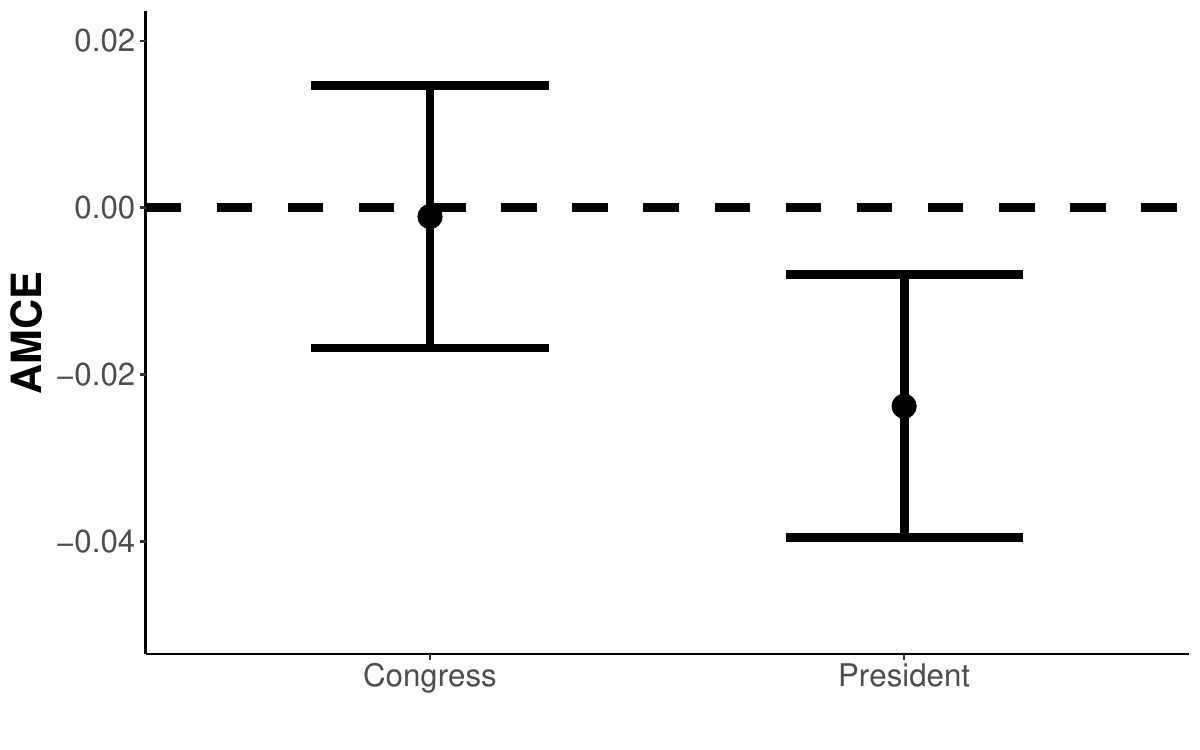"}
\caption{The estimated Average Marginal Component Effect (AMCE) of candidate's gender in the \cite{gender} study. We present the estimates for congressional candidates (left) and presidential candidates (right). The 95\% confidence intervals are also shown.}
\label{fig:gender_replication}
\end{center}
\end{figure}

The original study yields a negative estimate of the AMCE of female candidates, relative to male counterparts, for presidential candidates. However, the estimated AMCE of female candidates is not statistically distinguishable from zero for congressional candidates, based on a simple $t$-test for the coefficient of \textit{Male} from the linear regression with cluster standard errors. This finding led to the authors' conclusion that gender discrimination ``is limited to presidential rather than congressional elections'' (p. 583). Figure~\ref{fig:gender_replication} reproduces these AMCE estimates. Like the immigration example, the authors fit a linear regression with all fourteen factors as predictors to obtain these estimates. 

In this paper, we use the CRT to formally test whether the gender of congressional candidates matters in \textit{any way} for voters' preferences while controlling for the other candidate characteristics.  The rejection of this null hypothesis would indicate that gender does matter even for congressional candidates. 

\subsection{Limitations of Existing Approaches}
\label{subsection:AMCE_limitations}

Although the AMCE is a useful causal quantity of interest and can be easily and reliably estimated, it is not free of limitations.  The AMCE is a marginal effect based on two types of averaging: (1) averaging over the distribution of other attributes, and (2) averaging over the responses (and hence respondents).  Recall that in the standard causal inference setting with a binary treatment, a zero average treatment effect does not necessarily imply zero treatment effect for everyone.  The treatment may benefit some and harm others, and these positive and negative effects can cancel out through averaging.  The AMCE suffers from a similar problem, potentially masking important causal heterogeneity if there are interactions among attributes and/or between attributes and respondent characteristics. In the immigration conjoint experiment described above, for example, the overall AMCE estimates suggest little difference across countries of origin.  And yet \citeauthor{immigration} show that respondents with high ethnocentrism may have a certain preference over certain countries of origin when compared to those with low ethnocentrism. 

Additionally, although an AMCE-based confidence interval that does not contain zero represents evidence that the factor matters, a narrow AMCE-based confidence interval that contains zero only implies that a factor has a weak \textit{marginal} effect. Nevertheless, practitioners tend to use narrow AMCE-based confidence intervals that contain zero to conclude that a factor does not matter.  For example, \cite{AMCE} conclude that the ``candidates' income does not matter much'' and that the ``candidates' racial and ethnic backgrounds are even less influential'', based on the AMCE-based confidence interval for income and ethnicity (p. 19). Similarly, \citeauthor{gender} conclude that gender effects are ``limited to only Congressional candidates'' (p. 3), based on AMCE-based confidence interval of gender for Congressional candidate.

Although the AMCE is  popular in conjoint analysis, especially among political scientists, there also exist model-based approaches to flexibly estimate potentially any quantity of interest. In particular, logistic regression remain a popular model-based alternative in conjoint analysis \citep{mcfa:73, marketgeneral,marketreg} especially in marketing research.\footnote{Although a hierarchical modeling approach remains another popular model-based alternative in conjoint studies, we do not consider it here because it is based on a Bayesian framework rather than frequentist approach taken in this paper \citep{HB_conjoint}.} Despite the flexibility of logistic regression, model misspecification remains a significant challenge. Although researchers may add more interactions to account for all possible effects, such an approach can reduce statistical power and more importantly lead to invalid $p$-values \citep{phasetransition}. We show in Appendix~\ref{appendix:logistic_reg} Figure~\ref{fig:appendix_logistic} that using logistic regression and accounting for all two-way interactions to reduce model misspecification can easily lead to invalid $p$-values. 

Consequently, a consensus among researchers has emerged that flexible machine learning algorithms are necessary for capturing these causal interactions \citep{distributional_effects,dispute2,dispute3,gopl:imai:pash:22}.  Yet machine learning algorithms, despite their flexibility, cannot yield consistent estimates in high-dimensional settings without strong assumptions.  In addition, statistical inference in small samples remains a challenge \citep{hard_pvals, hard_pvals2,imai:li:21}. Our goal is to address these problems through an assumption-free approach based on the conditional randomization test. 

\section{The Proposed Methodology}
\label{section:methodology}

In this section we describe the proposed methodology based on the Conditional Randomization Test \citep[CRT;][]{CRT}.  We show how to apply the CRT to conjoint analysis in order to test whether a factor of interest matters, without making any assumptions. We discuss various test statistics that can be used with the CRT and several useful extensions for conjoint analysis. 

\subsection{Notation and Setup}
\label{subsection:notation}
For concreteness, we focus on the forced-choice conjoint design, under which a respondent is asked to choose one of two profiles.  Our methodology is general and can be extended to other designs.  Let $n$ be the total number of respondents.  As is often done in practice, suppose that each respondent evaluates $J$ pairs of profiles, yielding a total of $nJ$ responses (for notational simplicity, we assume the same number of evaluations for each respondent). 

We use $Y_{ij} \in \{0, 1\}$ to represent the binary outcome variable for evaluation $j$ by respondent $i$, which equals 1 for selecting the left profile and 0 for choosing the right  profile.  Although for convenience we use ``left'' and ``right'' to distinguish two profiles under each evaluation, the profiles do not necessarily have to be placed side by side on the actual survey platform. We use the following $nJ \times 1$ stacked vector representation for this outcome variable $\bY = [\bY_1; \bY_2; \ldots; \bY_n]$, where $\bY_i = [Y_{i1}; Y_{i2}; \ldots; Y_{iJ}]$ of dimension $J \times 1$ denotes the outcome variable for respondent $i$. We use $[a_1; a_2; \dots; a_n]$ to denote a vertical stacking of vectors or matrices $a_1, a_2, \dots, a_n$. We often observe some characteristics of the respondents, and we use $\bV_i$ to denote a $J \times r$-dimensional matrix of $r$ pre-treatment covariates for respondent $i$ that are repeated across $J$ rows.

Next, let $p$ represent the total number of attributes or factors used for each conjoint profile. We use a scalar $X_{ij \ell}^{L} \in \{1, 2, \dots, K_{\ell}\}$ to denote the value of the $\ell$th factor of interest for evaluation $j$ by respondent $i$, where the superscript distinguishes the factors for the left ($L$) and right ($R$) profiles, and $K_{\ell} \ge 2$ is the total number of factor levels for factor $\ell$. We use $\bX_{ij}^{L} = [X_{ij1}^L; \dots; X_{ijq}^L]$ to denote a $q$-dimensional column vector, containing all $q$ factors of interest for the left profile for respondent $i$ in the $j$th evaluation where $q \le p$. We define $\bX_{ij}^R$ similarly for the right profile. In addition, we use $\bX_{ij} =[\bX_{ij}^{L}; \bX_{ij}^{R}]$ as a column vector of length $2q$ to represent the main factors of interest from two profiles together. Lastly, the remaining $(p-q)$ factors are denoted by $\bZ_{ij} = [\bZ_{ij}^{L}; \bZ_{ij}^{R}]$, where each term is similarly defined. For example in the immigration conjoint experiment, if the main factor of interest is ``country of origin'', the other factors include ``education'' and ``profession''. 

As done for the outcome variable, we stack all evaluation-specific factors to define respondent-level factor matrices, which are further combined to yield the factor matrix $\bX=[\bX_{1}; \bX_{2}; \ldots;\bX_{n}]$ and $\bZ=[\bZ_{1}; \bZ_{2}; \ldots; \bZ_{n}]$ of dimension $nJ \times 2q$ and $nJ \times 2(p-q)$, respectively, where $\bX_i =[\bX_{i1}^\top; \bX_{i2}^\top; \ldots; \bX_{iJ}^\top]$ and $\bZ_i=[\bZ_{i1}^\top; \bZ_{i2}^\top; \ldots; \bZ_{iJ}^\top]$ are matricies of dimension $J \times 2q$ and $J \times 2(p-q)$, respectively. Lastly, we also stack all respondent characteristics $\bV = [\bV_1; \bV_2; \dots; \bV_n]$ of dimension $nJ \times r$.

Finally, we use $\bY(\bx, \bz)$ to denote the $nJ$-dimensional vector of the potential outcomes when $\bX = \bx$ and $\bZ = \bz$. This notation implies that we avoid assuming the no interference effect in the Stable Unit Treatment Value Assumption (SUTVA) since our vector of potential outcomes is a function of the entire set of treatments $\bX$ and $\bZ$ \citep{rubi:90}. We assume a super-population framework, where the potential outcomes $\bY(\bx, \bz)$ are assumed to be drawn from a population of infinite size. In Appendix~\ref{appendix:finitepop_inf},  we discuss how our framework is related to a finite-population framework, which is the basis of Fisher's randomization test. In conjoint analysis, the profile attributes are randomized according to a known distribution, $P(\bX, \bZ)$.  Our framework is general, allowing for any randomization distribution.  For example, some may randomize each factor independently using complete randomization whereas others may induce dependency among factors by removing a certain set of attribute combinations. In general, the randomization of the factors implies the following independence relation,
\begin{equation}
    \bY(\bx, \bz) \independent (\bX, \bZ) \hspace{0.1cm} \text{ for all } \bx \in \mathcal{X},\ {\rm and}\ \bz \in \mathcal{Z}, \label{eq:randomization}
\end{equation}
where we use $\mathcal{X}$ and $\mathcal{Z}$ to represent the support of $\bX$ and that of $\bZ$, respectively (see Chapter 3.6 of \citep{rubin:imbens}).

\subsection{The Conditional Randomization Test}
\label{subsection:CRT_intro}
The Conditional Randomization Test (CRT) is an assumption-free approach that combines design-based inference with flexible machine learning algorithms. For ease of presentation, we first introduce the CRT without incorporating the respondent characteristics $\bV$ and then return in Section~\ref{subsection:incorporating_respondent} to show how $\bV$ can be incorporated in all the proposed methods. The CRT allows us to examine whether the factors of interest $\bX$ change the response $\bY$ while holding the other factors $\bZ$ constant. Specifically, we test the following null hypothesis,
\begin{equation}
H_{0}: \bY(\bx,\bz) \overset{d}{=} \bY(\bx', \bz) \hspace{0.1cm} \text{ for all } \bx, \bx' \in \mathcal{X},\ {\rm and}\ \bz \in \mathcal{Z}, \label{eq:null}
\end{equation}
where we use $\overset{d}{=}$ to denote distributional equality. As a reminder, $H_0$ states that our entire \textit{vector} of potential outcomes are equal in distribution for any values of $\bX$. Our alternative hypothesis states that $\bX$ affects $\bY$ in some way while keeping $\bZ$ unchanged.  This is formalized as,
\begin{equation}
\quad H_{1}: \bY(\bx,\bz) \overset{d}{\neq} \bY(\bx',\bz) \hspace{0.1cm} \text{ for some } \bx, \bx' \in \mathcal{X},\ {\rm and}\ \bz \in \mathcal{Z}. \label{eq:alternative}
\end{equation}

We emphasize that the null hypothesis defined in Equation~\eqref{eq:null} implies the absence of any causal effects involving the main factor(s) of interest.  For example, the null hypothesis is false if $\bX$ affects $\bY$ for \emph{any} individual respondent or subgroup of respondents. Similarly, the null hypothesis does not hold if $\bX$ influences $\bY$ only when $\bZ$ takes a certain set of values.  Thus, the null hypothesis precludes any heterogeneous or interaction effects as well. 

Contrast this hypothesis test formulation with that of the standard AMCE-based analysis, which asks whether each factor of interest $\bX_{i}$ matters {\it on average}. More specifically, \citeauthor{AMCE} assume each individual's potential outcome is only a function of its own profile task, i.e., $Y_{ij}(\bX,\bZ) = Y_{ij}(\bX_{ij}, \bZ_{ij})$, and computes the marginal importance of $\bX_i$ by averaging each individual potential outcome over $\bZ_i$ as well as the respondents, which are assumed to be exchangeable, leading to the following null hypothesis,
\begin{equation}
 H_0^{\text{AMCE}}: \mathbb{E}\{Y_{ij}(\bx, \bZ_{ij})\} \ = \ \mathbb{E}\{Y_{ij}(\tilde{\bx}, \bZ_{ij})\}, 
\end{equation}
where $\bx$ and $\tilde{\bx}$ are the specified values of the main factors and the expectation is taken over $\bZ_{ij}$ (other factors) and the respondents. As briefly explained in Section~\ref{subsection:AMCE_limitations}, the limitation of the AMCE-based approach is that averaging over other factors can mask important causal interaction and heterogeneity.

We now establish the equivalence between the null hypothesis about the potential outcomes defined in Equation~\eqref{eq:null} and the conditional independence relation among observed variables.  This result allows us to use the CRT, which is a general assumption-free methodology for testing conditional independence relations in designed experiments \citep{CRT}.  We state this result as the following theorem whose proof is given in Appendix~\ref{subsection:associationcausation}.
\begin{theorem}[Equivalence] \label{th:equivalence}
The null hypothesis defined in Equation~\eqref{eq:null} is equivalent to the following conditional independence hypothesis under the randomization assumption of Equation~\eqref{eq:randomization}, 
$$H_0^{\text{CRT}}: \bY \independent \bX \mid \bZ.$$
\end{theorem}

The CRT is a general assumption-free methodology in designed experiments that produces exact $p$-values without asymptotic approximation.  The CRT combines the advantages of both design-based and model-based approaches by enabling the use of any test statistic, including ones based on complex machine learning (ML) algorithms, while making no modeling assumptions.  In the conjoint analysis literature, researchers have used traditional regression modeling \citep[e.g.,][]{logistic, clogitapplication, mcfa:73} and more recently modern ML algorithms \citep[e.g.,][]{egam:imai:19,dispute2}.  However, the validity of these analyses critically depends on modeling assumptions, parameter tuning, and/or asymptotic approximation. In contrast, the CRT assumes nothing about the conditional distribution of the outcome $\bY$ given $(\bX,\bZ)$.  Indeed, it does not even require the data to be independently or identically distributed, a property which we use later to test carryover and profile order effects.  The only requirement is the specification of the conditional distribution of $\bX$ given $\bZ$, which is readily available from the experimental design of conjoint analysis. Although the power of the CRT critically depends on the test statistic, the CRT always controls type 1 error no matter what the true model is. This contrasts with other model based approaches that require modeling assumptions to be valid (see Appendix~\ref{appendix:logistic_reg} for more details).

\begin{algorithm}[t]
 \textbf{Input:} Data $(\bX,\bY,\bZ)$, test statistic $T(\bx,\by,\bz)$, total number of re-samples $B$, conditional distribution $\bX \mid  \bZ$\;
  \For{$b = 1, 2, \dots, B$}{
  Sample $\bX^{(b)}$ from the distribution of $\bX \mid \bZ$\ conditionally independently of $\bX$ and $\bY$;
 }
 \textbf{Output:} $p$-value $:= \frac{1}{B+1} \left[1 + \sum_{b=1}^{B} \mathbbm{1}\{T(\bX^{(b)}, \bY, \bZ) \geq T(\bX, \bY, \bZ)\}\right]$\footnotemark
 \caption{Conditional Randomization Test (CRT)} \label{algo:CRT}
\end{algorithm}
\footnotetext{We add one to the numerator and denominator so that the distribution of the $p$-value is stochastically dominated by the uniform distribution as suggested by \citep{CRT}.}
Algorithm~\ref{algo:CRT} summarizes the general procedure used to compute the exact $p$-value for the CRT. Note that if $\bX$ and $\bZ$ are independently randomized, as is often the case, one can simply sample $\bX^{(b)}$ from the marginal distribution of $\bX$.  If, on the other hand, certain combinations of $\bX$ and $\bZ$ values (e.g., doctor without a college degree in the immigration conjoint experiment) are excluded, then we must use the appropriate conditional distribution of $\bX$ given $\bZ$. Critically, Algorithm~\ref{algo:CRT} is valid for complicated experimental designs, so long as one can sample from the conditional distribution $\bX$ given $\bZ$. The CRT can be computationally intensive since it requires computing the test statistic $T$ a total of $B+1$ times. However these computations can easily be parallelized. Furthermore, recent works \citep{HRT, dCRT} have shown that certain test statistic constructions also alleviate the need for these computations. In Appendix~\ref{appendix:dICRT}, we detail several tricks that can be used to dramatically reduce the computation time when implementing the CRT. For the main application results presented in Section~\ref{section:empirical_results} (first column of Table~\ref{tab:mainresults}), we note that the parallelized computational time was approximately six minutes with 50 cores to calculate each $p$-value with $B = 2,000$. Our software package makes it easy for practitioners to use multiple cores and provides a step-by-step instruction for using many cores on Amazon Web Services.\footnote{The detailed instructions and example use cases can be found in a vignette of our open-source software package in \url{https://cran.r-project.org/web/packages/CRTConjoint/vignettes/CRTConjoint.html}.}

The $p$-value of the CRT is valid\footnote{That is, under $H_0$, $P(p\text{-value}\le\alpha)\le\alpha$ for all $\alpha\in [0,1]$.} regardless of sample size and test statistic \citep{CRT}. To see this, it suffices to recognize that under the null hypothesis all $B+1$ test statistics, $T(\bX, \bY, \bZ)$, $T(\bX^{(1)}, \bY, \bZ)$, $\dots$, $T(\bX^{(B)}, \bY, \bZ)$, are exchangeable given $(\bY, \bZ)$. While any test statistic produces a valid $p$-value under the CRT, the choice of test statistic determines the statistical power.  We now turn to this practically important consideration.   

\subsection{Test Statistics}
\label{subsection:teststat}

To obtain a powerful test statistic that does not mask important interactions, we consider a test statistic based on the Lasso logistic regression with hierarchical interactions, or HierNet \citep{hiernet}. We note that if researchers wish to target main effects without considering interactions, the $F$-statistic from a standard linear regression of the response $\bY$ on $\bX$ is a reasonable test statistic to use. HierNet allows for the regularization of all possible two-way interaction terms while respecting their hierarchy.  Specifically, HierNet constrains the two-way interaction effects to be smaller in magnitude than their corresponding main effects.  For example, this implies that a two-way interaction effect will be set to zero if its relevant main effects are all zero. A stricter regularization on the interactions is desirable because the space of possible two-way interactions is large and grows quadratically, and we expect many to be indistinguishable from zero. Lastly, when fitting HierNet, we use the dummy variable encoding (i.e., each factor level is represented by its own dummy variable) but do not omit the baseline level.  We can fit this overparameterized model because of the regularization of HierNet.  The primary advantage of this approach is that the results are no longer dependent on the choice of baseline levels \citep{egam:imai:19}. 

We begin by considering the simplest case where we have a single main factor of interest $\bX$ ($q = 1$).  Without loss of generality, we assume that this is the first factor among the total of $p$ factors.  There are two types of interaction effects to consider \citep{distributional_effects}.  First, a within-profile interaction effect represents the interaction between one level of the main factor and another level of a different factor within the same profile.  Second, a between-profile interaction effect represents the factor interaction between two profiles (left versus right) that are being compared under the forced choice design.  

Our proposed test statistic is based on the sum of relevant squared main and interaction effects after subtracting their respective means, 
\begin{equation}
\begin{aligned}
T_{\text{HierNet}} & = \underbrace{\sum_{k=1}^{K_1} (\hat\beta_{k} - \bar\beta)^2}_{\text{main effects}} + \underbrace{\sum_{\ell = 2}^{p} \sum_{k=1}^{K_1} \sum_{k^\prime = 1}^{K_{\ell}} (\hat\gamma_{1 \ell kk^\prime} - \bar\gamma_{1 \ell k^\prime})^{2}}_{\text{within-profile interaction effects}} +  \underbrace{\sum_{\ell = 1}^{p} \sum_{k=1}^{K_1} \sum_{k^\prime = 1}^{K_{\ell}} (\hat\delta_{1 \ell kk^\prime} - \bar\delta_{1 \ell k^\prime})^{2}}_{\text{between-profile interaction effects}},
\end{aligned}
\label{eq:hiernet}
\end{equation}
where $\hat\beta_{k}$ is the estimated main effect coefficient for the $k$th level of our factor of interest $\bX$ with $\bar{\beta}$ denoting the average of these estimated main effect coefficients, and $\hat\gamma_{1 \ell kk^\prime}$ and $\hat\delta_{1 \ell kk^\prime}$ represent the estimated within-profile and between-profile interaction effect coefficients between the $k$th level of the factor of interest $\bX$ and the $k^\prime$th level of the $\ell$th factor, respectively.  Similar to the main effects, $\bar{\gamma}_{1 \ell k^\prime}$ and $\bar{\delta}_{1 \ell k^\prime}$ denote the averages of their corresponding estimated interaction effect coefficients. We do not consider third or higher order interactions because of the typical sample size in a conjoint experiment and a lack of powerful methods to detect such interactions. However, in Section~\ref{section:empirical_results} we illustrate how to incorporate third order interactions when prior substantive knowledge is available.

This test statistic can be easily generalized to the setting where there is more than one factor of interest ($q > 1$).  In such a case, we simply compute Equation~\eqref{eq:hiernet} for each factor of interest, and then sum the resulting values to arrive at the final test statistic. $T_{\text{HierNet}}$ aims to capture any differential effects the levels of $\bX$ have on the response through their main effects and relevant interaction effects. For example, suppose that $\bX$ is ``gender'' in the political candidate conjoint experiment (Section~\ref{subsection:gender}). Under $H_0$, we would expect all main effects and any interaction effects of \textit{male} and \textit{female} to be roughly equivalent, thus making $T_{\text{HierNet}}$ close to zero. However, suppose that \textit{male} candidates with a certain ``education level'' were favored more than \textit{female} candidates with a certain ``education level''. Then, we would expect these interactions to differ, making $T_{\text{HierNet}}$ further from zero.  

We use cross-validation\footnote{The CRT remains valid if used with cross-validation so long as the resampled test statistics based on $\bX^b$ similarly uses cross-validation.  This ensures exchangeability.} to choose the value of HierNet's tuning parameter, which controls the degree of regularization. In addition, when the sample size and the number of factors are large, fitting HierNet can be computationally demanding. To alleviate this issue, we propose computational speedups of the HierNet test statistics, which are detailed in Appendix~\ref{appendix:dICRT}. In particular, we drop $\bX$ when fitting the HierNet tuning parameter via cross-validation.  Since this computationally expensive step does not depend on $\bX$ , we do not need to re-run it for each $\bX^b$.

So far, we have constructed our test statistic as if there is no profile order effect. This implies that the effects of each factor do not depend on whether it belongs to the left or right profile.  Formally, we have imposed the following symmetry constraints in our HierNet test statistic, 
\begin{equation}
\begin{aligned}
    \hat\beta_{k} \ & = \ \hat\beta_{k}^{L} \ = \ - \hat\beta_{k}^{R}, \quad
    \hat\gamma_{1\ell kk^\prime} \  = \ \hat\gamma_{\ell \ell^\prime kk^\prime}^{L} \ = \ -\hat\gamma_{\ell \ell^\prime kk^\prime}^{R}, \quad \hat\delta_{\ell \ell^\prime kk^\prime} \ = \ -\hat\delta_{\ell^\prime \ell k^\prime k},
    \end{aligned}\label{eq:constraints}
\end{equation}
where the superscripts $L$ and $R$ denote the left and right profile effects, respectively. $\hat\delta_{\ell \ell^\prime k k^\prime}$ denotes the between profile interaction between the $k$th level of factor $\ell$ in the left profile with the $k^\prime$th level of factor $\ell^\prime$ in the right profile.\footnote{Equation~\eqref{eq:constraints} also implies that the between-profile interactions in Equation~\eqref{eq:hiernet} for the same factor obey $\hat \delta_{\ell \ell k k'} = -\hat \delta_{\ell \ell k' k}$ for any factor $\ell$ and levels $k, k'$. In particular this implies that $\hat \delta_{\ell \ell k k} = 0$, i.e., between-profile interactions of the same factor and same level are zero, while $\hat \delta_{\ell \ell k k'}$ are counted twice in Equation~\eqref{eq:hiernet}, i.e., between-profile interactions of the same factor and levels $k$ and $k'$ are counted twice in Equation~\eqref{eq:hiernet}.} The signs of the estimated coefficients reflect the fact that the response variable $\bY$ is recorded as 1 if the left profile is chosen and as 0 if the right profile is selected.  These constraints reduce the dimension of parameters to be estimated by half.  

Importantly, the validity of the proposed tests does not depend on whether the assumption of no profile order effect holds.   Through simulations, Figure~\ref{fig:constrained_vs_unconstrained} of Appendix~\ref{appendix:constraint_TS} shows that these constraints can significantly increase statistical power when there is no profile order effect. To incorporate this symmetry constraint, we append another copy of the dataset below the original data set, where the appended copy is identical to the original dataset except that the order of left and right profiles is flipped and the response variable is transformed as $\mathbf{1} - \bY$ before fitting HierNet (see Appendix~\ref{appendix:constraint_TS} for details). In Section~\ref{subsection:extensions}, we show how to use the CRT for testing the validity of the assumption of no profile order effect.

Because the validity of the CRT does not depend on modeling assumptions, one can incorporate a variety of assumptions into test statistics.  In general, test statistics have a greater statistical power if the assumptions hold in the true (unknown) data generating process.  Therefore, as much as possible the choice of test statistic should reflect researchers' substantive knowledge as illustrated in one of our empirical analyses (see Section~\ref{subsection:gender_results}). 

\subsection{Generalization of the Null Hypothesis and Test Statistic}
\label{subsection:generalization}

Researchers are often interested in testing only a few levels of interest as opposed to testing the whole factor. Yet, simply dropping the observations that correspond to those irrelevant factor levels can lead to a loss of statistical power. An advantage of the formulation described below is that we can retain all observations including those whose factor levels are irrelevant, which can improve statistical power. For example, suppose we are interested in how respondents differentiate immigrants from \textit{Mexico} and \textit{Germany}. If the way in which respondents differentiate between immigrants from \textit{Mexico} and those from \textit{Germany} is different from how they distinguish between immigrants from \textit{Mexico} and those from \textit{China}, then this implies that the respondents are viewing immigrants from \textit{Mexico} differently than those from \textit{Germany}. Therefore, detecting any differences for even the irrelevant levels may help improve the statistical power. 

Here, we generalize the null hypothesis and test statistic, given in Equations~\eqref{eq:null}~and~\eqref{eq:hiernet}, so that the methodology can accommodate any combinations of factor levels.  We introduce a coarsening function $h$ that groups factor levels of interest while assigning other factor levels to themselves.  Formally, this coarsening function is defined as $h: \mathcal{X} \mapsto \widetilde{\mathcal{X}}$, where $|\mathcal{X}| \ge |\widetilde{\mathcal{X}}|$. Thus, for our aforementioned immigration example, $h$ will assign the same value to immigrants from \textit{Mexico} and \textit{Germany} while leaving all other combinations mapped to different values. 

Under this setup, we can test the null hypothesis that specific levels within $\bX$ do not affect the potential outcome in any way.  Formally, 
\begin{align}
H_0^{\text{General}}: \bY(\bx,\bz) \overset{d}{=} \bY(\bx',\bz) \hspace{0.1cm} \text{for all} \ \bx, \bx' \in \mathcal{X}, \ \text{such that } h(\bx) = h(\bx')\ \text{and}\ \bz \in \mathcal{Z}. \label{eq:genH0}
\end{align}
The condition $h(\bx) = h(\bx')$ enables the comparison of the factor levels of interest alone. Additionally, $H_0$ is a special case of $H_0^{\text{General}}$ when the coarsening function $h$ is the identity function.  Finally, applying the same argument as the one used to prove Theorem~\ref{th:equivalence}, it can be shown that $H_0^{\text{General}}$ is equivalent to the following conditional independence relation,
\begin{equation}
    \bY \independent \bX \mid h(\bX), \bZ.
\label{eq:CRT_general}
\end{equation}

To test this null hypothesis, we first fit the same HierNet with the main effects and two-way interaction effects. To incorporate the coarsening function $h$, our test statistic takes the same form as the one given in Equation~\eqref{eq:hiernet} but is based only on the estimated coefficients that correspond to the factor levels of the group induced by the $h$ function, i.e., \textit{Mexico} and \textit{Germany} in the above example.  Appendix~\ref{appendix:differentialeffects} contains further details of testing the general null hypothesis and the corresponding CRT algorithm. In Section~\ref{subsection:immigration_results}, we also provide an example of applying $H_0^{\text{General}}$ that contains more details on this test statistic.  Under our framework, we do not need to drop observations that have irrelevant factor levels, thus increasing statistical power as mentioned above. Finally, Appendix~\ref{appendix:grouping_factor_levels} details  how to further generalize $H_0^{\text{General}}$ when a researcher is interested in grouping factor levels, i.e., combining levels \textit{France}, \textit{Germany}, and \textit{Poland} into one level \textit{Europe} when testing $H_0^{\text{General}}$. Although coarsening via $h$ also involved ``grouping'' levels, the grouping described in Appendix~\ref{appendix:grouping_factor_levels} aggregates factor levels to allow comparison between higher-level categories while the coarsening function $h$ allows us to focus our hypothesis test only on differences between a subset of factor levels of interest.

\subsection{Testing the Regularity Assumptions of Conjoint Analysis}
\label{subsection:extensions}

To further demonstrate the flexibility of the CRT, we also show how to use the CRT for testing the validity of several commonly made assumptions of conjoint analysis. Although we were interested in \textit{rejecting} $H_0$ above, we are interested in \textit{accepting} the null hypothesis for the hypotheses presented in this section. Therefore, we propose test statistics that are designed to be reasonably powerful for general settings in conjoint analysis.

\paragraph{Profile Order Effect.}
The assumption of no profile order effect states that changing the order of profiles, i.e., left versus right, does not affect the actual profile chosen (since the value of $\bY$ corresponds to whether the left or right profile is chosen, $\bY$ should be recoded as $\mathbf{1}-\bY$ when the profile order is changed). We denote the potential outcome $Y_{ij}(\bx_{ij}^L, \bx_{ij}^R, \bz_{ij}^L, \bz_{ij}^R)$, which is now a function of left and right profiles. Although not necessary, we assume here no interference between responses for notational clarity (see Appendix~\ref{appendix:regularityassumption} for the general case). Lastly, we use $\mathcal{X}_\text{ind}$ and $\mathcal{Z}_\text{ind}$ to denote the support of $[\bx_{ij}^L; \bx_{ij}^R]$ and that of $[\bz_{ij}^L; \bz_{ij}^R]$ respectively, representing the support of factors used in each individual's evaluation (hence ``$\text{ind}$'' in the subscript).

We formally state the assumption of no profile order effect as the following null hypothesis that reordering of the left and right profiles has no effect on the adjusted response: 
\begin{align*}
H_0^{\text{Order}}: Y_{ij}(\bx_{ij}^L, \bx_{ij}^R, \bz_{ij}^L, \bz_{ij}^R)  \stackrel{d}{=} 1 -  Y_{ij}(\bx_{ij}^R, \bx_{ij}^L, \bz_{ij}^R, \bz_{ij}^L), \hspace{0.1cm} \text{for all} \ i, j, [\bx_{ij}^L; \bx_{ij}^R] \in \mathcal{X}_\text{ind}, \ [\bz_{ij}^L; \bz_{ij}^R] \in \mathcal{Z}_\text{ind}.
\end{align*}
We modify the HierNet test statistic in Equation~\eqref{eq:hiernet} with the same HierNet fit on $(\bX,\bY,\bZ)$ but without enforcing the constraints in Equation~\eqref{eq:constraints} as, 
\begin{eqnarray*}
    T_{\text{HierNet}}^{\text{Order}}(\bX,\bY,\bZ) & = & \sum_{\ell=1}^p \sum_{k=1}^{K_\ell} \left(\hat\beta_{\ell k}^{L}+\hat\beta_{\ell k}^{R}\right)^{2} + \sum_{\ell=1}^p \sum_{\substack{\ell^\prime=1 \\ \ell^\prime \ne \ell}}^p \sum_{k=1}^{K_\ell} \sum_{k^\prime = 1}^{K_{\ell^\prime} } \left(\hat\gamma_{\ell\ell^\prime kk^\prime}^{L} + \hat\gamma_{\ell\ell^\prime kk^\prime}^{R} \right)^2 \nonumber\\
& &  + \sum_{\ell=1}^p \sum_{\substack{\ell^\prime=1\\ }}^p  \sum_{k=1}^{K_\ell} \sum_{k^\prime = 1}^{K_{\ell^\prime} }\left(\hat\delta_{\ell\ell^\prime kk^\prime}+ \hat\delta_{\ell^\prime \ell k^\prime k} \right)^2.
\end{eqnarray*}
Since the symmetry constraints given in Equation~\eqref{eq:constraints} must hold under $H_0^{\text{Order}}$, a large value of this test statistic indicates a potential violation of the null hypothesis. To conduct the CRT for testing $H_0^{\text{Order}}$, we resample and recompute our test statistics.  Appendix \ref{appendix:regularityassumption} provides details about the testing procedure.

\paragraph{Carryover Effect.} Researchers also often rely on the assumption of no carryover effect \citep{AMCE}.  The assumption states that the order of the $J$ evaluations each respondent performs has no effect on the outcomes.  This assumption is violated, e.g., if respondents use information from their previous evaluations when assessing a given pair of profiles. To test this carryover effect, we assume no interference across respondents but consider potential interference across evaluations within each respondent.  

Let $\bx_{i,1:(j-1)}$ represent all the profile attributes that were presented to respondent $i$ from the first evaluation to the $(j-1)$th evaluation.  Then, the potential outcome can be written as a function of both current and previous profiles, i.e., $Y_{ij}\left(\bx_{i,1:(j-1)}, \bz_{i,1:(j-1)}, \bx_{ij}, \bz_{ij}\right)$ for $j \geq 2$, where we assume no interference between respondents but allow intereference within a respondent. Our null hypothesis is that, for a given evaluation $j \geq 2$, the response $Y_{ij}$ is independent of all the previous profiles conditional on the current profiles:
\begin{equation*}
    H_0^{\text{Carryover}}: Y_{ij}\left(\bx_{i,1:(j-1)}, \bz_{i,1:(j-1)}, \bx_{ij}, \bz_{ij}\right) \stackrel{d}{=} Y_{ij}\left(\bx_{i,1:(j-1)}', \bz_{i,1:(j-1)}', \bx_{ij}, \bz_{ij}\right),
\end{equation*}
where for all $\ i\geq 1,\ j  \geq 2,\ \bx_{i,1:(j-1)}, \bx_{i,1:(j-1)}' \in \mathcal{X}_\text{ind}^{j-1}, \ \bz_{i,1:(j-1)}, \bz_{i,1:(j-1)}' \in \mathcal{Z}_\text{ind}^{j-1}, \ \bx_{ij} \in \mathcal{X}_\text{ind}, \ \bz_{ij} \in \mathcal{Z}_\text{ind}$ with $\mathcal{X}_\text{ind}^{j-1}$ and $\mathcal{Z}_\text{ind}^{j-1}$ denoting the support of $\bx_{i,1:(j-1)}$ and that of $\bz_{i,1:(j-1)}$, respectively. 

We test this null hypothesis by using a test statistic that targets whether the immediately preceding evaluation affects the current evaluation. We believe targeting the lag-1 effect in the test statistic is reasonable because if a carryover effect exists, respondents are likely to be affected most by the immediately preceding evaluation. For example, if respondents believe that they have placed too much weight on profiles' professions in the previous evaluation, they might decide to rely on the current profiles' ``country of origin'' factor more than its ``profession'' factor in order to balance across evaluations.  Under this scenario, we would expect a significant interaction between previous profiles' ``profession'' factor and current profiles' ``country of origin'' factor.

We modify the test statistic given in Equation~\eqref{eq:hiernet} in the following way. Suppose that $J$ is even (if $J$ is odd, simply consider $J-1$ evaluations).  We first define a new response vector $\bY_i^\ast = [Y_{i2}; Y_{i4}; \ldots; Y_{iJ}]$ by taking every other evaluation.  Similarly, we can define new factors of interest $\bX_i^\ast= [[\bX_{i1}; \bZ_{i1}]^\top; [\bX_{i3}; \bZ_{i3}]^\top, \allowbreak\ldots; [\bX_{i,J-1}; \bZ_{i,J-1}]^\top]$ and a new set of conditioning variables $\bZ_i^\ast= [[\bX_{i2}; \bZ_{i2}]^\top; [\bX_{i4}; \bZ_{i4}]^\top, \allowbreak\ldots; [\bX_{iJ}; \bZ_{iJ}]^\top]$. We then fit HierNet with the new response $\bY^\ast = [\bY_1^\ast; \bY_2^\ast; \dots; \bY_n^\ast]$ on $(\bX^\ast,\bZ^\ast)$, where $\bX^\ast = [\bX_1^\ast; \bX_2^\ast; \dots; \bX_n^\ast]$ and $\bZ^\ast$ is defined similarly.

For this particular scenario, HierNet will estimate all main effects and interaction effects of $\bZ^\ast$, and the interaction effects between $\bX^\ast$ and $\bZ^\ast$, which are of primary interest. To increase the power of the test, we set all main and interaction effects of $\bX^\ast$ to zero since we do not expect the previous profile alone to impact the respondent's choice. Furthermore, we also do not expect the interaction effects between $\bX^\ast$ and $\bZ^\ast$ to differ, depending on the ordering (left versus right) of the relevant factors. Therefore, we enforce all these interaction effects to have equal magnitude as done in Equation~\eqref{eq:constraints} (see Appendix~\ref{appendix:constraint_TS} for further details). This leads to the following test statistic,
\begin{equation*}
T_{\text{HierNet}}^{\text{Carryover}}(\bX^\ast,\bY^\ast,\bZ^\ast) \ = \ \sum_{\ell=1}^p \sum_{\ell^\prime=1}^p \sum_{k=1}^{K_\ell} \sum_{k^\prime = 1}^{K_{\ell^\prime} } \hat\gamma_{\ell\ell^\prime kk^\prime}^{2},
\end{equation*}
where $\gamma_{\ell\ell^\prime kk^\prime}$ represents the coefficient of an interaction term between the $k$th level of the $\ell$th factor of the profile used in the previous evaluation and the $k^\prime$th level of the $\ell^\prime$th factor of the profile used in the current evaluation.  Appendix~\ref{appendix:regularityassumption} explains how to resample the test statistic in this setting.

\paragraph{Fatigue Effect.} Researchers may be concerned that a respondent performing a large number of conjoint evaluations may experience the ``fatigue effect,'' resulting in a declining quality of responses.  Recently, \citet{choice_task} conducted an empirical study, in which they examine how the pattern of responses depends on the number of evaluations each respondent performs.  While a typical conjoint experiment asks each respondent to carry out 5 to 10 evaluations, the authors increase this number up to 30 evaluations.  They find that the results are robust to increasing the number of evaluations. Here, we show how to use the CRT to formally test the presence of the fatigue effect. 

Similar to the carryover effect, we investigate whether there is a fatigue effect within each respondent's potential outcome $Y_{ij}(\bx_{ij},\bz_{ij})$, where we again assume no interference effect as done when testing no profile order effect. We test the following null hypothesis that the potential outcome is unaffected if respondent $i$ evaluated the same pair of profiles $(\bx_{ij}, \bz_{ij})$ but at a later or earlier evaluation $j' \neq j$:
\begin{equation*}
H_0^{\text{Fatigue}}: Y_{ij}(\bx_{ij}, \bz_{ij})  \stackrel{d}{=} Y_{ij'}(\bx_{ij}, \bz_{ij})  \ \text{for all} \ i, j, j', \ \bx_{ij} \in \mathcal{X}_\text{ind},\ \text{and}\ \bz_{ij} \in \mathcal{Z}_\text{ind}.
\end{equation*}

We propose a similar HierNet test statistic that reflects a scenario where respondents will only pay attention to a shrinking number of factors as they rate more profiles. In this case, we would expect interactions between the factors and the evaluation order index $\bF=(F_1,F_2,\ldots,F_n)$, which represents an $nJ$-dimensional integer vector with $\bF_i=(1,2,\ldots,J)$ for all $i=1,2,\ldots,n$. Again, for the sake of statistical power, we impose the absence of profile order effects on HierNet as done in Equation~\eqref{eq:hiernet}. Our proposed test statistic is the following from a HierNet fit of $\bY$ on $(\bX, \bZ, \bF)$,
\begin{equation*}
T_{\text{HierNet}}^{\text{Fatigue}}(\bX,\bY,\bZ,\bF) \ = \ \sum_{\ell=1}^p \sum_{k=1}^{K_\ell}  \hat\gamma_{\ell k }^{2},
\end{equation*}
where $\hat\gamma_{\ell k}$ represents the coefficient of an interaction term between $\bF$ and level $k$ of factor $\ell$. Appendix \ref{appendix:regularityassumption} shows how to resample and recompute the test statistics to test $H_0^{\text{Fatigue}}$.

\subsection{Incorporating Respondent Characteristics}
\label{subsection:incorporating_respondent}

In conjoint experiments, researchers often expect factors of interest to interact strongly with respondent characteristics \citep{gender, immigration, immigration_debate}. It is possible to exploit this fact when applying the CRT by directly incorporating respondent characteristics, $\bV$, into the test statistic. Doing so can substantially increase the statistical power.

We incorporate respondent characteristics in the CRT procedure by appending $\bV$ to $\bZ$ and holding both $(\bZ,\bV)$ constant.  Since respondent characteristics are not randomized factors, unlike $\bZ$, $\bV$ is not guaranteed to be independent of the potential outcomes.  We can test the following causal null hypothesis that conditions upon $\bV$,
\begin{equation}
H_{0}: \bY(\bx,\bz) \overset{d}{=} \bY(\bx',\bz) \mid \bV \hspace{0.1cm} \text{ for all } \bx, \bx' \in \mathcal{X},\ {\rm and}\ \bz \in \mathcal{Z}.
\label{eq:null_with_V}
\end{equation}
Theorem~\ref{th:equivalence} can be easily extended to show that this null hypothesis is equivalent to the conditional independence relation $\bY \independent \bX \mid \bZ, \bV$. Algorithm~\ref{algo:CRT} also stays the same except we sample $\bX^{b}$ from the distribution of $\bX \mid (\bZ, \bV)$ and our test statistic is now a function of $(\bX, \bY, \bZ, \bV)$. 

A major benefit of incorporating respondent characteristics is the ability to capture respondent characteristic interactions into the test statistic. Consequently, we incorporate an additional predictor $\bV$ when fitting HierNet and modify the HierNet test statistic in Equation~\eqref{eq:hiernet} as,
\begin{equation} 
\begin{aligned}
T_{\text{HierNet}}  =&  \underbrace{\sum_{k=1}^{K_1} (\hat\beta_{k} - \bar\beta)^2}_{\text{main effects}} + \underbrace{\sum_{\ell = 2}^{p} \sum_{k=1}^{K_1} \sum_{k^\prime = 1}^{K_{\ell}} (\hat\gamma_{1 \ell kk^\prime} - \bar\gamma_{1 \ell k^\prime})^{2}}_{\text{within-profile interaction effects}} \\
&+\underbrace{\sum_{\ell = 1}^{p} \sum_{k=1}^{K_1} \sum_{k^\prime = 1}^{K_{\ell}}(\hat\delta_{1 \ell kk^\prime} - \bar\delta_{1 \ell k^\prime})^{2}}_{\text{between-profile interaction effects}} + \sum_{m=1}^r \sum_{k=1}^{K_1} \sum_{w = 1}^{L_{m} } \underbrace{(\hat\xi_{1 m k w} - \bar\xi_{1 m w})^{2}}_{\text{respondent interaction effects}},
\end{aligned}
\label{eq:modified_hiernet}
\end{equation}
where $\hat\xi_{1 m k w}$ represents the interaction between the $k$th level of our factor of interest $\bX$ and the $w$th level of the $m$th factor of a respondent characteristic. If a respondent characteristic is a numeric variable, we could either coarsen it into a factor variable or directly include it in the model as a numeric variable.\footnote{HierNet standardizes all variables when performing the fit. Therefore, even if a numeric variable is on a different scale, all estimated coefficients remain comparable.} Again, $\bar{\xi}_{1 m w}$ denotes the average of these estimated interaction coefficients. Similar to Equation~\eqref{eq:modified_hiernet}, we add the additional constraint that $\hat\xi_{\ell m k w} = \hat\xi_{\ell m k w}^{L}  = \hat\xi_{\ell m k w}^{R}$, where the superscript $L$ and $R$ similarly denotes the left and right profile effects. Finally, we modify $T^{\text{Order}}_{\text{HierNet}}$ similarly by adding $\sum_{\ell=1}^p \sum_{m=1}^r \sum_{k=1}^{K_\ell} \sum_{w = 1}^{L_{m} } (\hat\xi_{\ell m k w}^{L} + \hat\xi_{\ell m k w}^{R}  )^{2}$ to the original test statistic in $T^{\text{Order}}_{\text{HierNet}}$ to account for respondent characteristic interactions. For all empirical applications in Section~\ref{section:empirical_results}, we incorporate $\bV$ and use the modified test statistics to test $H_0$ and the no profile order effect. 

\section{Simulation Studies}
\label{section:simulations}

A primary advantage of the CRT is that it can yield powerful statistical tests by incorporating machine learning algorithms to capture complex interactions in high dimensions.  The CRT achieves this while maintaining the finite sample validity of the resulting $p$-values.  In this section, we conduct simulation studies to show that the CRT with the HierNet test statistic can be substantially more powerful than the AMCE-based test.  

For simplicity, our simulation setting assumes one evaluation for each respondent ($J = 1$). We show additional simulations with multiple evaluations with respondent random effects in Figure~\ref{fig:RE_sim} of Appendix~\ref{appendix:multiple_tasks_sim}. In addition, there is only one main factor of interest $\bX$ ($q = 1$) and ten other factors $\bZ$ and no $\bV$. Each factor is assumed to have two levels and is independently and uniformly randomized. To clearly separate main and interaction effects, we use the sum-to-zero constraint by coding each binary factor as $(-0.5, 0.5)$. Our response model, $\Pr(Y_i =1 \mid X_i, \bZ_i)$, follows a logistic regression that includes main effects for $(\bX,\bZ)$ and within-profile and between-profile interaction effects between $\bX$ and $\bZ$ and among $\bZ$. Our setup also has no profile order effects.

Our goal is to examine the difference in statistical power between the CRT and AMCE-based tests in the presence of interactions. Thus, we vary the size of the interaction effect as well as the number of non-zero interaction effects between $\bX$ and $\bZ$. Throughout all of our simulations, we use an effect size of $0.1$ for all main effects for $(\bX,\bZ)$ and an effect size of $0.05$ for all interactions among $\bZ$ and fix the sample size to $n = 3,000$. We also make all non-zero interactions between $\bX$ and $\bZ$ equal in size for all simulations.  As shown in Appendix~\ref{appendix:sim_heterogenous}, the power of the CRT remains identical if there are heterogeneous interaction effects with varying sizes. Appendix~\ref{appendix:simulation} provides further details of our simulation setup.

For each simulated data set, we test the null hypothesis of no causal effect using the CRT and AMCE-based approaches. We then compute the proportion of times each test rejects the null hypothesis. For the CRT, we use the HierNet test statistic given in Equation~\eqref{eq:hiernet} obtained by fitting the response $\bY$ on our predictors $(\bX, \bZ)$ and enforcing the constraints shown in Equation~\eqref{eq:constraints}.  For the AMCE, we use the $t$-test based on the estimated regression coefficient of $\bX$ obtained through fitting the response $\bY$ on a single predictor $\bX$.\footnote{Although this is a valid procedure to compute the AMCE estimate for $\bX$, practitioners typically compute the AMCE estimates of all factors ($\bX, \bZ$) simultaneously with a single linear regression of $\bY$ on all $(\bX, \bZ)$.  Figure~\ref{fig:AMCE_long} of Appendix~\ref{appendix:AMCE_sim} shows that the power of the AMCE remains indistinguishable when using all factors $(\bX, \bZ)$ in a single linear regression. Lastly, although our goal is to compare the CRT with the AMCE, we acknowledge that practitioners may also use the omnibus $F$-test for testing interactions by including all the two-way interactions. We show in Appendix~\ref{appendix:logistic_reg} that such an approach leads to inflated $p$-values, thus we omit this as a baseline comparison here.}

\begin{figure}[t]
\begin{center}
\includegraphics[width=\textwidth]{"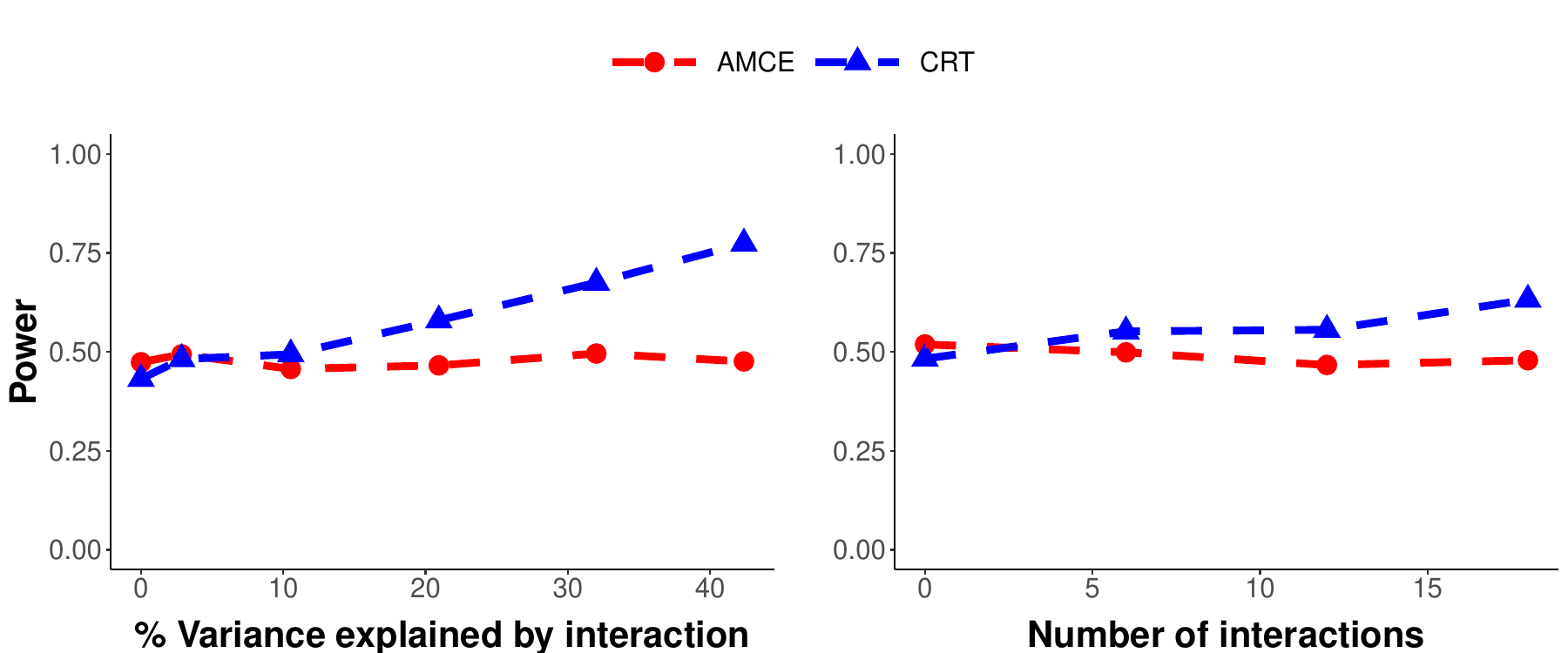"}
\caption{The figure shows how the power of the CRT and AMCE-based tests varies as the size of interaction effects (left plot) or the number of non-zero interaction effects (right) increases. The AMCE-based test (red circles) is based on the $t$-test from the estimated regression coefficient.  The CRT uses the HierNet test statistic given in Equation~\eqref{eq:hiernet}. The sample size is $n = 3,000$. Finally, the standard errors are negligible with a maximum value of $0.016$.}
\label{fig:interaction}
\end{center}
\end{figure}

The left plot of Figure~\ref{fig:interaction} shows how the statistical power of each test varies as the size of interaction effects between $\bX$ and $\bZ$ increases.  The number of non-zero interaction effects between $\bX$ and $\bZ$ is fixed at six.  For ease of interpretation, we plot the percentage of total outcome variance explained by the interaction effects between $\bX$ and $\bZ$ on the $x$-axis.\footnote{The $x$-axis ticks correspond to interaction sizes of $0, 0.025, 0.05, 0.075, 0.1$, and $0.125$, respectively. For example, the 20\% point on the $x$-axis refers to an interaction size of $0.05$.} In our setup, the total variance represents the outcome variance explained by all main and interaction effects under the latent representation of the logistic regression model (see Appendix~\ref{appendix:simulation} for details). Consistent with our theoretical expectation, the CRT (blue triangles) becomes more powerful than the AMCE-based test (red circles) as the interaction size increases. For example, when the interaction size is strong enough to account for about 30\% of the total variance, the CRT is approximately 20 percentage points more powerful than the AMCE-based test. When there is no interaction effect, the CRT is only slightly less powerful (by about 3 percentage points) than the AMCE-based test. 

The right plot of Figure~\ref{fig:interaction} shows how the power of the tests change as one varies the number of non-zero interaction effects.  The size of interaction effects is fixed to $0.06$, around half the size of the main effect. We find that as expected, the CRT becomes more powerful than the AMCE-based test as the number of interaction increases. For example, when there are twelve interactions the CRT is approximately 10 percentage points more powerful than the AMCE-based test. Even when there is no interaction effect at all, the loss of statistical power is minimal. Appendix~\ref{appendix:constraint_TS} presents additional simulation results, showing that the use of no profile order constraints given in Equation~\eqref{eq:constraints} increases the power of test. 

\section{Application to Conjoint Experiments}
\label{section:empirical_results}

In this section, we apply the proposed CRT to the two conjoint studies introduced in Section~\ref{section:empirical_examples}.  

\subsection{Immigration Preferences and Ethnocentrism}
\label{subsection:immigration_results}

We begin our analysis of the immigration conjoint experiment by testing whether respondents differentiate between immigrants from Mexico and those from European countries. We use the same data set as the one used in \citet{immigration}.  This gives us a total sample of $6,980$ observations with $n = 1,396$ respondents each rating $J = 5$ tasks. Our main factor of interest $\bX$ is the ``country of origin'' variable. Since we are only interested in testing how respondents differentiate Mexican and European candidates, we use the generalized hypothesis $H_0^{\text{General}}$ defined in Equation~\eqref{eq:genH0} and coarsen the three levels --- \textit{Germany}, \textit{France}, and \textit{Poland} --- into one level called \textit{Europe} (see Appendix~\ref{appendix:grouping_factor_levels} for a formal treatment).  Furthermore, the $h$ function in $H_0^{\text{General}}$ takes the ``country of origin'' variable and maps the relevant levels of \textit{Mexico} and \textit{Europe} to one output and the remaining levels to other unique outputs. We include all the other randomized factors and respondent characteristics (see Section \ref{subsection:immigration}) as $\bZ$ and $\bV$, respectively, except the ethnocentrism variable, which is only measured for a subset of respondents. We incorporate this variable at the end of this section.

We fit HierNet using $\bY$ as the response and our main factor $\bX$, other randomized factors $\bZ$, and respondent characteristics $\bV$ as the predictors.  We then compute the test statistic given in Equation~\eqref{eq:modified_hiernet} while imposing the implied no profile order effect constraints given in Equation~\eqref{eq:constraints} (with the constraints applied to the respondent characteristic interactions too).  As mentioned briefly in Section~\ref{subsection:generalization}, we slightly modify this test statistic by only using the estimated coefficients for \textit{Mexico} and \textit{Europe} while ignoring the other coefficients. 

\begin{table}[!t]
\begin{center}
\begin{adjustbox}{max width=\textwidth,center}
\begin{tabular}{lcc|ccc} 
 & CRT & AMCE & Profile order effect & Carryover effect & Fatigue effect  \\ 
Immigration & 0.042 & 0.27 & 0.80 & 0.12 & 0.45 \\
Gender & 0.026 & 0.93, 0.40 & 0.15 & 0.97 & 0.66 \\ 
\end{tabular}
\end{adjustbox}
\caption{The $p$-values based on the Conditional Randomization Test (CRT) and the Average Marginal Component Effect (AMCE) Estimation. The two rows represent the different applications. The first two columns present the $p$-values from the HierNet-based CRT and AMCE-based test statistics.  The first row presents the $p$-values for testing whether the immigrant's ``country of origin'' (\textit{Mexico} or \textit{Europe}) matters for immigration preferences while the second row tests whether candidate's ``gender'' matters for voters' preferences of Congressional candidates, respectively. The second AMCE-based $p$-value for ``gender'' corresponds to a fair comparison with the CRT-based $p$-value by additionally testing the ``gender'' interaction with the candidate's ``party affiliation'' (\textit{Democratic} or \textit{Republican}). The remaining columns report the $p$-values for testing no profile order effect, no carryover effect, and no fatigue effect for the respective application.}
\label{tab:mainresults}
\end{center}
\end{table}

As shown in the left upper cell of Table~\ref{tab:mainresults}, the CRT $p$-value of this test statistic is 0.042, providing evidence that respondents differentiate immigrants from \textit{Mexico} and \textit{Europe}. For comparison, we also compute the $p$-value based on the estimated AMCE of being from \textit{Mexico} compared to being from \textit{Europe}. We apply a commonly used linear regression approach described in Section~\ref{subsection:immigration} to compute this $p$-value. Specifically, we first fit a linear regression model using ``country of origin'', ``reason of immigration'', and their interaction as predictors to account for the restricted randomization \citep{AMCE}. The standard errors are clustered by respondent. We then use the $F$-test of the linear equality constraint that implies the null hypothesis under the linear model. As shown in the upper cell of the second column of Table~\ref{tab:mainresults}, the resulting $p$-value for the difference between {\it Mexico} and {\it Europe} is 0.27, which is statistically insignificant.

The above result suggests that the CRT may be able to capture complex interactions and yield greater statistical power than the AMCE-based test.  The two largest interactions in the observed test statistic are within-profile interactions between ``country of origin'' and ``education'' and between ``country of origin'' and ``prior trips to U.S.'' factors, which included whether or not the immigrant entered the U.S. illegally.  Thus, we next assess the degree to which the interaction effects account for this difference in statistical power.  To do this, we use a Lasso logistic regression without interaction terms where we only include the main effects of $(\bX, \bZ, \bV)$. The CRT $p$-value of using only the relevant levels of the main effects of $\bX$ as the test statistic is $0.082$, which is somewhat larger than the $p$-value based on HierNet test statistic. This suggests that interactions play some role in yielding a more powerful test than the AMCE-based approach. 

\citeauthor{immigration} suggest that respondents do not differentiate between immigrants from \textit{Mexico} and those from \textit{Europe} based on the results of their main analysis (see Figure~\ref{fig:immigration_prior_results}, which replicates this analysis).  However, they also conduct a subgroup analysis and find that ``country of origin'' has statistically significant interaction(s) with the respondent's ethnocentrism through a subgroup analysis \citep[see also][for related findings]{immigration_debate}.  Thus, we now repeat the same analysis as above except that we include this ethnocentrism variable as an additional respondent characteristic in $\bV$.  Note that unlike the original analysis, we do not dichotomize this variable and use the original continuous scale. Since ethnocentrism is only measured for white and black respondents, the number of total respondents is reduced to $n = 1,135$. Despite this reduction in sample size, the inclusion of the ethnocentrism variable produces the $p$-value of 0.019, which is smaller than the $p$-value of the analysis without this variable. As expected, the largest interaction in the observed test statistic involves the ethnocentrism variable.  All together, our analysis provides evidence that respondents differentiate immigrants from \textit{Mexico} and \textit{Europe}.

Lastly, we use the CRT to test the the three commonly made regularity assumptions of conjoint analysis: no profile order effect, no carryover effect, and no fatigue effect. The last three columns in Table~\ref{tab:mainresults} present the $p$-values from the various tests described in Section~\ref{subsection:extensions}. We find no evidence that these assumptions are violated in the immigration conjoint experiment (the first row). In particular, the fact that we do not detect profile order effects suggests that imposing the symmetry constraint as done in Equation~\eqref{eq:constraints} likely improves power. 

\subsection{Role of Gender in Candidate Evaluation}
\label{subsection:gender_results}

For the gender conjoint experiment, we test whether or not the gender of Congressional candidates matters in voter preferences.  We use the same data as the one used in \citet{gender}. The Congressional dataset consists of 7,915 observations with 5 tasks performed by each of $n = 1,583$ respondents. Our main factor of interest $\bX$ is a binary variable representing \textit{male} or \textit{female}.  In addition, we use the remaining 12 randomized factors $\bZ$ and all the respondent characteristics $\bV$ (see Section \ref{subsection:gender}). We test the main null hypothesis $H_0$ introduced in Section \ref{subsection:CRT_intro}.  

As mentioned in Section \ref{subsection:teststat}, the use of substantive knowledge can improve the power of the test. To demonstrate this, we leverage the Presidential candidate dataset from the same conjoint experiment to find the strongest interaction with the gender of candidates.  We then include this interaction term as an additional main effect in HierNet when computing the test statistic given in Equation~\eqref{eq:modified_hiernet}.\footnote{Our test statistic $T_{\text{HierNet}}(\bX, \bY, \bZ,\bV)$ is not only a function of the Congressional dataset $(\bX, \bY, \bZ, \bV)$ but also the entire presidential dataset ($\mathbf{P}$). Therefore, we must now hold all $(\bZ,\bV, \mathbf{P})$ fixed in the resampling procedure. However, this does not change Algorithm \ref{algo:CRT} and the resulting $p$-value remains valid because $\bX \mid (\bZ,\bV, \mathbf{P})$ is still an independent fair coin flip between the levels of \textit{male} and \textit{female}.} By including it as a main effect, HierNet applies less shrinkage on this interaction term. In addition, HierNet will consider potential three-way interactions involving this interaction term and other variables in $\bZ$.  The power will be greater if strong interactions in the Presidential candidate data are also present in the Congressional candidate data.

To find the strongest interaction in the Presidential candidate dataset, we obtain a CRT $p$-value for each variable in $(\bZ,\bV)$ with a test statistic that focuses on the interaction strength for the corresponding variable under consideration. Specifically, the test statistic uses Lasso logistic regression with all main effects of $(\bX,\bZ,\bV)$ and an additional interaction between $\bX$ and one variable from $(\bZ,\bV)$ (see Appendix~\ref{appendix:presidential_data} for more details). We choose the variable with the lowest $p$-value as the strongest interaction. The Presidential data shows that the candidate's ``party affiliation'' (\textit{Democratic} or \textit{Republican}) had the most significant interaction with their ``gender''.  Appendix~\ref{appendix:presidential_data} contains further details and a robustness check by repeating the same analysis but choosing the variable with the second lowest $p$-value as the additional main effect. 

As shown in the second row of Table \ref{tab:mainresults}, the CRT $p$-value using the HierNet test statistic is 0.029, showing that gender may matter even for Congressional candidates.  We find the largest two interactions in the observed test statistic were two three-way interactions: one between ``gender'', ``party affiliation'', and ``respondent's political interest'' and the other between ``gender'', ``party affiliation'', and the ``respondent's party affiliation''. This result is consistent with the findings in \citet{distributional_effects}, which suggests the existence of higher order interactions involving party affiliation. We assess the role of interaction effects using the same procedure above in the immigration example. The resulting CRT $p$-value from a Lasso logistic regression with only the main effects of $(\bX,\bZ,\bV)$ and the additional interaction with ``gender'' and ``party affiliation'' is $0.15$, suggesting that the other interactions were also helpful in detecting significance. 

For comparison, we compute the $p$-value based on the estimated AMCE of ``gender'' for Congressional candidates as presented in Figure~\ref{fig:gender_replication}. Similar to the common strategy used for the immigration conjoint experiment, we fit a linear regression model using ``gender'' as the sole predictor while clustering standard errors by respondents. We find that the $p$-value is 0.89. However, since the CRT leveraged the Presidential candidate data to up-weight the interaction with ``party affiliation'', we also create a fair comparison by obtaining analogous $p$-values for the AMCE. To do this, we again fit a linear regression using ``gender'' but with an additional main effect of ``party affiliation'' and the interaction of ``gender'' and ``party affiliation''. We then report the $p$-value from a $F$-test for both the main effect of ``gender'' and the interaction with ``party affiliation''. The resulting $p$-value is 0.40, showing that the AMCE-based result remains statistically insignificant. Finally, the last three columns in the second row show no evidence that the regularity assumptions are violated for this conjoint experiment.\footnote{We only test the no profile order effect assumption for Congressional candidates because this is the data relevant to the research question. However, for testing the carryover effect and fatigue effect, we use the full dataset including the Presidential candidates in order to increase power.}

\section{Concluding Remarks}
\label{section:conclusion}

Conjoint analysis is a popular methodology for analyzing multi-dimensional preferences and decision making. In this paper, we propose an assumption-free approach for conjoint analysis based on the conditional randomization test (CRT).  The proposed methodology allows researchers to test whether a set of factors of interest matter at all without assuming a statistical model. We also extend the proposed methodology to test for differential effects for any combination of factor levels and other regularity assumptions commonly invoked in conjoint analysis like the profile order effect.  Unlike the standard AMCE analysis, the CRT can avoid masking important interactions due to averaging over other factors.  When constructing CRT test statistics, researchers can use machine learning algorithms and/or domain knowledge to detect complex interactions among factors, without making modeling assumptions. The CRT is easy to implement and provides exact (i.e., non-asymptotic) $p$-values that are valid even in high dimensions.  We believe that this flexibility combined with its assumption-free nature makes the CRT a powerful tool for conjoint analysis. The CRT can complement the existing methods like the AMCE analysis by providing a useful way to examine whether a factor of interest matters at all.

\newpage
\bibliography{bibtex_file,my,imai} 
\bibliographystyle{pa}

\newpage
\appendix

\section{Proof of Theorem~\ref{th:equivalence}}
\label{subsection:associationcausation}

\begin{proof}
We first prove that if $\bY \independent \bX \mid \bZ$, then $\bY(\bx, \bz) \overset{d}{=} \bY(\bx', \bz)$ for any $\bx$, $\bx' \in \mathcal{X}$ and $\bz \in \mathcal{Z}$. 
\begin{align*}
   P(\bY \mid \bZ = \bz) &= P(\bY \mid \bX = \bx, \bZ = \bz)  \\
   &= P(\bY(\bx, \bz) \mid \bX = \bx, \bZ = \bz)  \\
   &= P(\bY(\bx, \bz) \mid \bZ = \bz) \\
   &= P(\bY(\bx,\bz)),
\end{align*}
where the third and fourth equalities follow from the randomization of $\bX$ and that of $\bZ$, respectively.  Similarly, we can show $P(\bY \mid \bZ = \bz) = P(\bY(\bx', \bz))$, thus we have shown $P(\bY(\bx, \bz)) = P(\bY(\bx',\bz))$ for any $\bx$, $\bx' \in \mathcal{X}$ and $\bz \in \mathcal{Z}$.

To the prove the other direction, we want to show that $\bY(\bx, \bz) \overset{d}{=} \bY(\bx', \bz)$ implies $\bY \independent \bX \mid \bZ$, or equivalently $P(\bY \mid \bX = \bx, \bZ= \bz) = P(\bY \mid \bX = \bx', \bZ = \bz)$ for any value of $\bz \in \mathcal{Z}$ and any value of $\bx$, $\bx' \in \mathcal{X}$. 
\begin{align*}
   P(\bY(\bx,\bz)) &= P(\bY(\bx,\bz) \mid \bZ = \bz) \\
   &= P(\bY(\bx,\bz) \mid \bX = \bx, \bZ= \bz)  \\
   &= P(\bY \mid \bX = \bx, \bZ = \bz),
\end{align*} 
where the first two equalities follow from the randomization of $\bZ$ and that of $\bX$, respectively.  The same argument shows $P(\bY(\bx',\bz)) = P(\bY \mid \bX = \bx', \bZ = \bz)$.
Finally, because $\bY(\bx, \bz) \overset{d}{=} \bY(\bx', \bz)$ we have that $P(\bY(\bx,\bz)) = P(\bY(\bx',\bz))$, implying $P(\bY \mid \bX = \bx, \bZ= \bz) = P(\bY \mid \bX = \bx', \bZ =\bz)$ for any value of $\bz \in \mathcal{Z}$ and any value of $\bx$, $\bx' \in \mathcal{X}$.
\end{proof}

\section{Relation to Finite-Population Inference}
\label{appendix:finitepop_inf}
In Section~\ref{section:methodology}, we introduced $H_0$ under the super-population framework. Here, we consider the relationship between $H_0$ and Fisher's sharp null of no treatment effect \citep{fisher:1935}.  Under the finite-population framework where the potential outcomes are fixed and the randomness comes only from the randomization of treatment assignment, if we assume no interference between units, $H_0$ reduces to testing $Y_{ij}(\bx_{ij}, \bz_{ij}) = Y_{ij}(\bx'_{ij}, \bz_{ij})$ for all $i,j$ and all possible values of $\bx_{ij}, \bx_{ij}' \in \mathcal{X}$ and $\bz_{ij} \in \mathcal{Z}$. Under the super-population framework, if we make the same no-interference assumption, $H_0$ reduces to testing $Y_{ij}(\bx_{ij}, \bz_{ij}) \overset{d}{=} Y_{ij}(\bx'_{ij}, \bz_{ij})$ for all $i,j$ and all possible values of $\bx_{ij}, \bx_{ij}', \bz_{ij}$, where the potential outcomes are assumed to be drawn from a population. These two null hypotheses are not equivalent even though the CRT can test $H_0$ under both the finite-population and super-population frameworks (as proven in Theorem~\ref{th:equivalence}).  This is because the distribution of $\bY(\bx, \bz)$ can still be equal to $\bY(\bx', \bz)$ even if $Y_{ij}(\bx, \bz)$ is different than $Y_{ij}(\bx', \bz)$ for some $i,j$. 

\section{Grouping Factor Levels}
\label{appendix:grouping_factor_levels}

In this appendix, we further detail how to test $H_0^{\text{General}}$ when the analyst is interested in grouping multiple factor levels. For example, in the immigration conjoint application (Section~\ref{subsection:immigration_results}), we wish to test whether respondents differentiate immigrants from Mexico and those from Europe where there are three distinct levels for countries from Europe: \textit{Germany}, \textit{France}, and \textit{Poland}. We now formally show how to group these levels up into one category \textit{Europe} and test the hypothesis $H_0^{\text{General}}$. 

We introduce a coarsening function $c$ that takes $q$ factors of interest $\bX$ as input and transforms them to a new set of grouped factors $\overline{\bX}$. Formally, this function is defined as $c: \mathcal{X} \mapsto \overline{\mathcal{X}}$ where $\overline{\mathcal{X}}$ represents the support of the grouped factors $\overline{\bX}$ with $|\mathcal{X}| \ge |\overline{\mathcal{X}}|$. For the above example, we define the function $c$ on the ``country of origin'' factors such that the \textit{Mexico} level takes one value whereas the \textit{France}, \textit{Germany}, and \textit{Poland} levels all take another value: \textit{Europe} $\in \overline{\bX}$.  All other factor levels are mapped to different values in $\overline{\bX}$.

Next, we define the outcome for each level of newly transformed factor levels.  Given the coarsening function $c$ defined above, we introduce the marginalized potential outcome variable $\overline{\bY}(\overline{\bx}, \bz)$, which averages over the distribution of original factor levels that are grouped.  Formally, this new outcome variable has the following mixture structure,
\begin{equation}
\overline{\bY}(\overline{\bx}, \bz) \ = \ \frac{\sum_{\bx' \in \mathcal{X}} \mathbf{1}\{c(\bx') = \overline{\bx}\}\bY(\bx',\bz)P(\bX = \bx' \mid \bZ = \bz)}{\sum_{\bx' \in \mathcal{X}} \mathbf{1}\{c(\bx') = \overline{\bx}\}P(\bX = \bx' \mid \bZ = \bz)},
\end{equation}
where $\bz \in \mathcal{Z}$, $\overline{\bx} \in \overline{\mathcal{X}}$, and $P(\bX = \bx \mid \bZ = \bz)$ represents the conditional distribution of $\bX$ given $\bZ$ used in the experiment.  For example, if we group three European countries---\textit{France}, \textit{Germany}, and \textit{Poland}---and create one new factor level \textit{Europe}, then its marginalized potential outcome will be a mixture distribution of the original potential outcomes for the three countries weighted by their known randomization probabilities conditional on the other factors. 

Furthermore, the previously introduced coarsening function $h$ now takes the newly grouped up factor $\overline{\bX}$ and maps it to the new coarsened factor, i.e., $h: \overline{\mathcal{X}} \mapsto \widetilde{\mathcal{X}}$. Consequently, our updated generalized null hypothesis is,
\begin{align}
\overline{H}_0^\text{General}: \overline{\bY}(\overline{\bx}, \bz)  \stackrel{d}{=} \overline{\bY}(\overline{\bx}', \bz) \hspace{0.1cm} \text{for all} \ \overline{\bx}, \overline{\bx}' \in \overline{\mathcal{X}}, \ \text{such that } h(\overline{\bx}) = h(\overline{\bx}')\ \text{and}\ \bz \in \mathcal{Z}. \label{eq:genH02}
\end{align}
Finally, it can also be shown by applying the same argument as the one used to prove Theorem~\ref{th:equivalence} that $\overline{H}_0^\text{General}$ is equivalent to the following conditional independence relation,
\begin{equation}
    \overline{\bY} \independent \overline{\bX} \mid h(\overline{\bX}), \bZ.
\label{eq:CRT_general2}
\end{equation}

To test this null hypothesis, we keep the original test statistic $T_{\text{HierNet}}$ shown in Equation~\eqref{eq:hiernet} under the same symmetry constraints given in Equation~\eqref{eq:constraints} except that we use $\overline{\bX}$ in place of $\bX$ to account for coarsening based on the function $c$. 

\newpage
\section{Data Description}
\label{appendix:data_table}

Tables~\ref{tab:immigration_factorlevel}~and~\ref{tab:gender_factorlevel} present all the factors and their respective levels used in the immigration conjoint experiment (Section~\ref{subsection:immigration}) and the gender candidate conjoint experiment (Section~\ref{subsection:gender}), respectively. 

\begin{table}[h!]
\footnotesize
\centering
\begin{tabularx}{\linewidth}{|c|L|} 
\hline
\textbf{Factor for Immigration Conjoint Experiment}  & \textbf{Factor Levels} \vspace{0.3cm} \\ 
\hline
Education level  &  
      No Formal Education, 
      Fourth Grade, 
      Eight Grade, 
      High School, 
      Two Years College, 
      College Degree, 
      Graduate Degree
    \\ 
\hline
Gender  &  
      Female, 
      Male 
   \\
\hline
Country of origin  &   
      Germany, 
      France, 
      Mexico, 
      Philippines, 
      Poland, 
      India, 
      China,  
      Sudan, 
      Somalia, 
      Iraq 
   \\
\hline
Language & 
      Fluent English, 
      Broken English, 
      Tried to speak English but unable to, 
      Spoke through an interpreter
   \\
\hline
Reason for application  & 
      Reunite with family members, 
      Seek better job, 
      Escape political/religious persecution 
   \\
\hline
Profession  &  
      Gardener, 
      Waiter, 
      Nurse,  
      Teacher,
      Child Care Provider, 
      Janitor, 
      Construction Worker, 
      Financial Analyst,  
      Research Scientist, 
      Doctor, 
      Computer Programmer 
   \\
\hline
Job experience  & 
      No job training or prior experience, 
      One to two years, 
      Three to five years,  
      More than five years 
   \\
\hline
Employment plan  & 
      Has a contract with a U.S. employer, 
      Does not have a contract with a U.S. employer, but has done job interviews, 
      Will look for work after arriving in the U.S., 
      Has no plans to look for work at this time 
   \\
\hline
Prior trips to the U.S. &  
      Never been to the U.S., 
      Entered the U.S. once before on a tourist visa, 
      Entered the U.S. once before without legal authorization, 
      Has visited the U.S. many times before on tourist visas, 
      Spent six months with family members in the U.S. 
   \\
\hline
\end{tabularx}
\caption{All nine randomized factors and their respective levels in the conjoint experiment used in \cite{immigration}.}
\label{tab:immigration_factorlevel}
\end{table}

\begin{table}
\small
\centering
\begin{tabularx}{\linewidth}{|c|L|} 
\hline
\textbf{Factor For Gender Conjoint Experiment}  & \textbf{Factor Levels} \vspace{0.3cm} \\ 
\hline
Gender  &  
      Male, 
      Female
    \\ 
\hline
Age  &  
      36, 
      44, 
      52, 
      60, 
      68, 
      76
   \\
\hline
Race/ethnicity  & 
      White, 
      Black, 
      Hispanic, 
      Asian American
   \\
\hline
Family &   
      Single (never married), 
      Single (divorced), 
      Married (no child), 
      Married (two children)
   \\
\hline
Experience in public office  &  
      12 years, 
      8 years, 
      4 years, 
      No experience
   \\
\hline
Salient personal characteristics  & 
      Strong leadership, 
      Really cares about people like you, 
      Honest, 
      Knowledgeable, 
      Compassionate, 
      Intelligent
   \\
\hline
Party affiliation &  
      Republican, 
      Democrat
   \\
\hline
Policy area of expertise &  
      Foreign policy, 
      Public safety (crime), 
      Economic policy, 
      Health care, 
      Education, 
      Environmental issues
     \\
\hline
Position on national security  &  
      Cut military budget and keep U.S. out of war, 
      Maintain strong defense and increase U.S. influence
    \\ 
\hline
Position on immigrants  & 
      Favors guest worker program, 
      Opposes guest worker program
   \\
\hline
Position on abortion & 
      Pro-choice, 
      Pro-life, 
      Neutral
   \\
\hline
Position on government deficit &   
      Reduce through tax increase, 
      Reduce through spending cuts, 
      Does not want to reduce
   \\
\hline
Favorability rating among public &  
      34\%, 
      43\%, 
      52\%, 
      61\%, 
      70\%
   \\
\hline
\end{tabularx}
\caption{All thirteen randomized factors and their respective levels in the conjoint experiment used in \cite{gender}.}
\label{tab:gender_factorlevel}
\end{table}

\newpage

\section{Enforcing The No Profile Order Effect}
\label{appendix:constraint_TS}
We detail here a way to enforce the no profile order effect constraints in Equation~\eqref{eq:constraints} for general test statistics. We show through simulations that these constraints, when they hold, can substantially improve statistical power.

\begin{table}[t]
\begin{center}
\small
\begin{tabular}{ |c|c|c|c|c|c|c|} 
\hline
\textbf{Row Number}&\textbf{Gender$^{L}$} & \textbf{Gender$^{R}$ }& \textbf{Party$^{L}$}  & \textbf{Party$^{R}$} & \textbf{Respondent Age} & $\bY$\\ 
\hline
 1 & Male  & Female  & Democrat & Republican & 27& 1\\ 
\hline
\vdots  &\vdots  & \vdots  & \vdots & \vdots &  \vdots & \vdots\\ 
\hline
$nJ +1$& Female & Male &  Republican & Democrat & 27 & 0\\ 
\hline
\vdots  &\vdots  & \vdots  & \vdots & \vdots & \vdots & \vdots\\ 
\hline
\end{tabular}
\caption{Visual example of $D^{c}$ data matrix with $\bX$ as gender, $\bZ$ as party affiliation, and $\bV$ as the respondent's age. With a slight abuse of notation Gender$^{L}$ denotes the left profile's gender values. Gender$^{R}$, Party$^{L}$, and Party$^{R}$ are defined similarly.}
\label{tab:example_M}
\end{center}
\end{table}

Under these constraints, switching the "left" and "right" profile order does not change the value of the test statistic. We now formalize this intuition. First denote $D \in \mathbb{R}^{nJ \times (2p + r + 1)}$ as the regression data matrix composed of $(\bX,\bZ,\bV, \bY)$. Let $s_{E}: \mathbb{R}^{nJ \times (2p + r + 1)} \to \mathbb{R}^{nJ \times (2p + r + 1)} $ denote a function that takes a data matrix as an input and swaps the ``left'' and ``right'' profile order for rows $E$ of the data matrix, where $E \subset \{1, 2, \dots, nJ\}$. For example, suppose we swap just the first row and we denote $(\tilde{\bX}, \tilde{\bZ}, \tilde{\bV}, \tilde{\bY})$ as the columns for the output $s_{\{1\}}(D)$. Then $\tilde{\bX}_{11}^{L} = \bX_{11}^{R}$, $\tilde{\bX}_{11}^{R} = \bX_{11}^{L}$, $\tilde{\bZ}_{11}^{L} = \bZ_{11}^{R}$, $\tilde{\bZ}_{11}^{R} = \bZ_{11}^{L}$, and $\tilde{Y}_{11} = 1-Y_{11}$ and all remaining rows remain identical as the original data $D$. The function applies no swap to the respondent characteristic $\bV$ ($\tilde{\bV} = \bV$). 

We introduce a new data matrix $D^{c}$ that appends the original data matrix with a data matrix that swaps the profile order for all rows. Formally, $D^{c} = [D; s_{\{1, 2, \dots, nJ\}}(D)]$. Table~\ref{tab:example_M} shows an example of $D^{c}$ with $\bX$ as gender, $\bZ$ as party affiliation, and $\bV$ as the respondent's age. Conceptually, $D^{c}$ aims to destroy all information about the profile order since the ``left'' and ``right'' profile are now indistinguishable, thus ensuring any test statistic that uses $D^{c}$ will respect the no profile order effect constraints. The following lemma formally states this result.
\begin{lemma}
\label{lemma:constraint}
Let $T(\cdot)$ be a row invariant test statistic,\footnote{More formally we say $T(D)$ is row invariant if $T(D) = T(\pi(D))$ for any possible $\pi$, where $\pi$ denotes a possible permutation of the rows of $D$.} then for any $E \subset \{1, 2, \dots, nJ\}$ we have that $T(D^{c}) = T(s_{E}(D)^c)$, where $s_{E}(D)^c = [s_{E}(D); s_{1, 2, \dots, nJ}(s_{E}(D))]$.
\end{lemma}
\begin{proof}
The lemma holds because $D^{c} = s_{E}(D)^c$ up to row permutations. Using the assumption that $T(\cdot)$ is a row invariant test statistic we obtain the equality. Many algorithms like Random Forest have random initializations in the process so $T(D^{c}) = T(s_{E}(D)^c)$ may only hold up to distributional equality. 
\end{proof}

Lemma \ref{lemma:constraint} states that the test statistic will remain invariant to any relabelling of ``left'' and ``right'' profiles as long the test statistic is a function of the $D^c$ data matrix. All regression based algorithms will respect exact row invariance. Lemma~\ref{lemma:constraint} allows practitioners to build any test statistic, even ones that do not have natural ``left'' and ``right'' coefficients, while still enforcing the no profile order effect. In particular, we enforce the constraints in Equation~\eqref{eq:constraints} by using $D^{c}$ as the input in HierNet.

\paragraph{Simulations.}
We now show that imposing the no profile order effect constraints in Equation~\eqref{eq:constraints} can substantially increase statistical power under the same simulation setup in Section~\ref{section:simulations} when the no profile order effect assumption is satisfied. To evaluate the power gain, we also fit HierNet without imposing the constraints in Equation~\eqref{eq:constraints} by using the original data $D$.  We use $T_{\text{HierNet}}(\bX,\bY,\bZ)^L + T_{\text{HierNet}}(\bX,\bY,\bZ)^R$ as the test statistic, where $T_{\text{HierNet}}(\bX,\bY,\bZ)^L$ is the same as $T_{\text{HierNet}}(\bX,\bY,\bZ)$ in Equation~\eqref{eq:hiernet} but all coefficients correspond to their respective estimates for the left profile. We similarly define $T_{\text{HierNet}}(\bX,\bY,\bZ)^R$. 

\begin{figure}
\begin{center}
\includegraphics[width=\textwidth]{"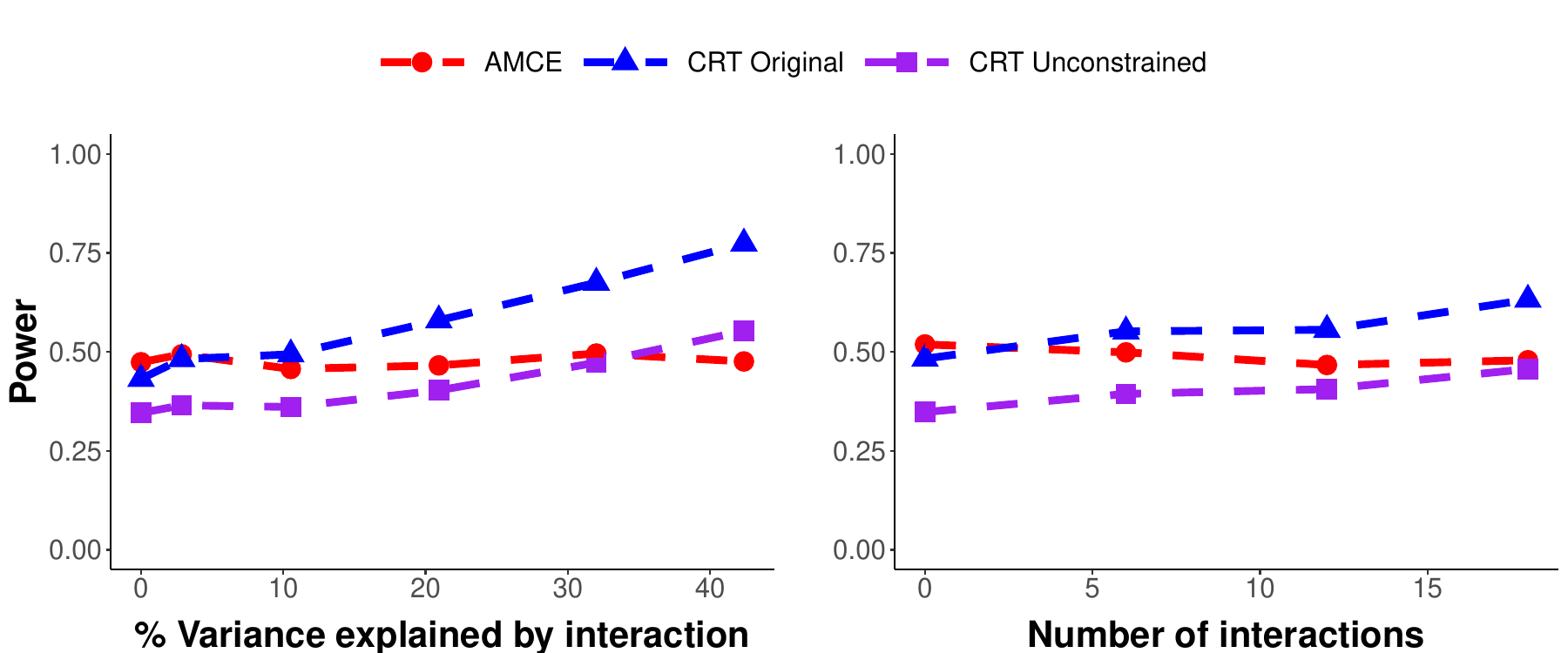"}
\caption{This figure represents the power gain from imposing the no ``Profile Order Effect'' constraints in Equation~\eqref{eq:constraints} under the same simulation setup in Figure~\ref{fig:interaction}. We keep the AMCE and the original HierNet statistical power curves (red circles and blue triangles, respectively) and add the new unconstrained HierNet test statistic (purple squares).}
\label{fig:constrained_vs_unconstrained}
\end{center}
\end{figure}

Figure~\ref{fig:constrained_vs_unconstrained} shows that imposing the no profile order effect constraints can significantly increase power. For example, the power of HierNet without the constraints (purple squares) is roughly equal or smaller than that of even the AMCE (red circles) when the interaction effect accounts for 20\% of the variance or when there are as many interactions as twelve. Furthermore, we see the power of using HierNet that imposes the constraints (blue triangles) is consistently higher than that of the HierNet without the constraints.

\paragraph{Enforcing No Profile Order Effect in the Carryover Effect Test Statistic.}

As mentioned in Section \ref{subsection:extensions}, we enforce the no profile order effect constraints when conducting the test of no carryover effect. We detail here how to implement this. 

Since $\bX^\ast$ represents the lag-1 profile values, we do not expect $\bX^\ast$ alone (without interactions with $\bZ^\ast$) to influence respondents' choice of the left versus right profile. Therefore, we force all main effects and interactions among $\bX^\ast$ to be zero in the following way. Let $\bX_i^{\ast L} \in\mathbb{R}^{\frac{J}{2} \times p}$ denote the columns of $\bX_i^\ast$ that correspond to the left profile in $\bX_i^\ast$. More formally, $\bX_i^{\ast L} = [[(\bX_{i1}^{L}); (\bZ_{i1}^{L})]^\top; \allowbreak [(\bX_{i3}^{L}); (\bZ_{i3}^{L})]^{\top}; \dots; [(\bX_{i,J-1}^{L}); (\bZ_{i, J-1}^{L})]^{\top}]$. If $J$ is not even, consider up to $J - 1$ instead. We define $\bX_i^{\ast R}, \bZ_i^{\ast L}, \bZ_i^{\ast R}$ similarly. We also define $\bX^{\ast L} = [\bX_1^{\ast L}; \dots; \bX_n^{\ast L}] \in\mathbb{R}^{\frac{nJ}{2} \times p}$ with $\bX^{\ast R}, \bZ^{\ast, L}, \bZ^{\ast, R}$ defined similarly. Then, we append copies of the following:
$[[(\bX^{\ast R})^\top; (\bX^{\ast L})^\top; (\bZ^{\ast R})^\top; (\bZ^{\ast L})^\top]^\top; \allowbreak [(\bX^{\ast L})^\top; (\bX^{\ast R})^\top; (\bZ^{\ast R})^\top; (\bZ^{\ast L})^\top]^\top; [(\bX^{\ast R})^\top; (\bX^{\ast L})^\top; (\bZ^{\ast L})^\top; (\bZ^{\ast R})^\top]^\top$
to the original data matrix 

\noindent $[(\bX^{\ast L})^\top; (\bX^{\ast R})^\top; (\bZ^{\ast L})^\top; (\bZ^{\ast R})^\top]^\top$, resulting in a total of $2nJ$ rows. Lastly, we also append copies of $[\mathbf{1} - \bY_i^\ast; \mathbf{1} - \bY_i^\ast; \bY_i^\ast]$ to the original response $\bY_i^\ast$. Appending the first copy of $[(\bX^{\ast R})^\top; (\bX^{\ast L})^\top; (\bZ^{\ast R})^\top; \allowbreak (\bZ^{\ast L})^\top]^\top$ to the original data matrix enforces the familiar constraint in Equation~\eqref{eq:constraints} using Lemma~\ref{lemma:constraint}. The remaining copies force all the main effects and interactions among $\bX^\ast$ to be zero by appealing to the same reasoning in Lemma~\ref{lemma:constraint}.

\section{CRT Procedure for Testing Extensions}
\label{appendix:extensions}

In this appendix, we describe in further detail how to carry out all the resampling procedures for the tests introduced in Sections~\ref{subsection:generalization}~and~\ref{subsection:extensions}.

\subsection{Testing $H_0^{\text{General}}$}
\label{appendix:differentialeffects}

When testing $H_0^{\text{General}}$, the resampling procedure is different than Algorithm~\ref{algo:CRT} because Equation~\eqref{eq:CRT_general} forces us to hold $(h(\bX), \bZ)$
constant rather than holding only $\bZ$ constant. The conditional distribution of $\bX \mid (h(\bX), \bZ)$ constrains $\bX$ to be randomized only within the factor levels of interest. For concreteness, consider the immigration example in Section~\ref{subsection:immigration_results}. In this case, $h(\bX)$ groups levels \textit{Mexico} and \textit{Europe} to one output while keeping all the other countries of origin the same. Therefore, when obtaining the resamples for the ``country of origin'' factor, we keep all countries except \textit{Mexico} and \textit{Europe} constant, i.e., countries such as \textit{China} do not get re-randomized. For the entries corresponding to \textit{Mexico} and \textit{Europe}, we resample values \textit{Mexico} and \textit{Europe} with probabilities $(0.25, 0.75)$, respectively (since 3 countries make up Europe). We present this resampling procedure in Algorithm~\ref{algo:CRT_generlized}. Lastly, we remark that Algorithm~\ref{algo:CRT_generlized} remains the same when testing $\overline{H}_0^{\text{General}}$ except we repalce $\bX$ with $\overline{\bX}$.

\begin{algorithm}[t]
 \textbf{Input:} Data $(\bX,\bY,\bZ, \bV)$, test statistic $T(\bx,\by,\bz, \bv)$, $h(\bx)$, total number of re-samples $B$\;
  \For{$b = 1, 2, \dots, B$}{
  Sample $\bX^{(b)}$ from the distribution of $\bX \mid (h(\bX), \bZ, \bV)$ conditionally independently of $\bX$ and $\bY$;
 }
 \textbf{Output:} $p$-value $:= \frac{1}{B+1} \left[1 + \sum_{b=1}^{B} \mathbbm{1}\{T(\bX^{(b)},\bY,\bZ, \bV) \geq T(\bX,\bY,\bZ, \bV)\}\right]$
 \caption{Generalized $H_0$: Testing $H_0^{\text{General}}$} \label{algo:CRT_generlized}
\end{algorithm}

\subsection{Testing the Regularity Assumptions}
\label{appendix:regularityassumption}

\paragraph{Profile Order Effect.}
We first describe testing the assumption of no profile order effect without imposing SUTVA. For any $E \subset\{1,\dots,nJ\}$, let $\bx^{(E)}$ swap the left and right profile values in rows $E$ of $\bx$ while leaving the remaining rows unchanged.  We similarly define $\bz^{(E)}$. Then, let $\bY^{(E)}(\bx, \bz)$ flip the bits of (replace 1's with 0's and vice versa) the entries of $\bY(\bx, \bz)$ corresponding to indices in $E$ while leaving the remaining entries unchanged. For example, for simplicity, assume no $\bZ$ while $nJ = 3$, $E = \{1 \}$, $\bx = [(G, F); (P, G); (M, F)]$ (the left profile values come first), and $\bY(\bx) = (1, 0, 1)$. Then $\bx^{(E)} = [(F, G); (P, G); (M, F)]$ and $\bY^{(E)}(\bx) = (0, 0, 1)$. The observed $\bY^{(E)}$ is defined similarly. We can formally state the null hypothesis of no profile order effect as follows:
\begin{align*}
H_0^{\text{Order}}: \bY(\bx, \bz) \ \stackrel{d}{=} \ \bY^{(E)}(\bx^{(E)}, \bz^{(E)}) \hspace{0.1cm} \text{for all} \ E \subset\{1,\dots,nJ\}, \bx \in \mathcal{X},\ \text{and}\ \bz \in \mathcal{Z}.
\end{align*}
This null hypothesis states that for all possible reorderings of the left and right profiles there is no causal impact on which profile is chosen. For the resampling procedure, we hold the realized values of all $(\bX,\bZ)$ constant while only resampling the subset $E$, i.e., drawing $nJ$ independent Bernoulli coin flips to determine which of the $nJ$ rows to include as part of $E$ as described above. Algorithm~\ref{algo:CRT_profileordereffect} details the procedure to calculate the CRT $p$-value for testing $H_0^{\text{Order}}$. Lastly, to not enforce Equation~\eqref{eq:constraints} in $T_{\text{HierNet}}^{\text{Order}}$, we fit HierNet on the original data matrix $(\bX,\bY,\bZ)$ rather than $D^{c}$.

\begin{algorithm}[t]
 \textbf{Input:} Data $(\bX,\bY,\bZ)$, test statistic $T(\bx, \by, \bz)$, total number of re-samples $B$\;
  \For{$b = 1, 2, \dots, B$}{
  Sample $nJ$ independent Bernoulli(0.5) independently of $(\bX,\bY,\bZ)$. $E^b$ is the index set corresponding to values of 1 in the $nJ$ Bernoulli's;
 }
 \textbf{Output:} $p$-value $:= \frac{1}{B+1} \left[1 + \sum_{b=1}^{B} \mathbbm{1}\{T(\bX^{(E^b)}, \bY^{(E^b)}, \bZ^{(E^b)}) \geq T(\bX, \bY, \bZ)\}\right]$
 \caption{CRT: Profile Order Effect} \label{algo:CRT_profileordereffect}
\end{algorithm}

\paragraph{Carryover Effect.}
When testing the carryover effect, we need to hold the even numbered tasks $\bZ^\ast$ fixed while resampling all odd numbered tasks $\bX^\ast$. Therefore, we resample all factors $(\bX_{ij}, \bZ_{ij})$ for $j = 1, 3, \dots, J-1$ from the experimental distribution for all the factors while holding $(\bX_{ij}, \bZ_{ij})$ for $j = 2, 4, \dots, J$ constant. Algorithm~\ref{algo:CRT_carryovereffect} details the procedure to calculate the CRT $p$-value for testing $H_0^{\text{Carryover}}$.

\begin{algorithm}[t]    
 \textbf{Input:} Data $(\bX^\ast, \bY^\ast, \bZ^\ast)$, test statistic $T(\bx^\ast, \by^\ast, \bz^\ast)$, total number of re-samples $B$\;
  \For{$b = 1, 2, \dots, B$}{
  Sample $(\bX_{ij}^{b}, \bZ_{ij}^{b})$ from the distribution of $(\bX_{ij}, \bZ_{ij})$ independently of $\bY$ for $i = 1, 2, \dots, n$ and $j = 1, 3, \dots J-1$. Let $(\bX_i^b)^\ast = [[\bX_{i1}^{b}; \bZ_{i1}^{b}]^\top; [\bX_{i3}^{b}; \bZ_{i3}^{b}]^\top; \allowbreak\ldots; [\bX_{i,J-1}^b; \bZ_{i,J-1}^b]^\top]$ and $(\bX^b)^\ast = [(\bX_1^\ast)^b; (\bX_2^\ast)^b; \dots; (\bX_n^\ast)^b]$;
 }
 \textbf{Output:} $p$-value $:= \frac{1}{B+1} \left[1 + \sum_{b=1}^{B} \mathbbm{1}\{T((\bX^{b})^\ast,\bY^\ast,\bZ^\ast) \geq T(\bX^\ast,\bY^\ast,\bZ^\ast)\}\right]$
 \caption{CRT: Carryover Effects} \label{algo:CRT_carryovereffect}
\end{algorithm}

\paragraph{Fatigue Effect.}
To carry out the CRT to test the fatigue effect, we re-sample only the task evaluation order index $\bF$ for each respondent uniformly from the set of all permutations on $\{1,\dots,J\}$, denoted as $\Pi_J$, while holding all the experimental factor values fixed. Algorithm~\ref{algo:CRT_fatigue} details the procedure to calculate the CRT $p$-value for testing $H_0^{\text{Fatigue}}$.

\begin{algorithm}[t]
 \textbf{Input:} Data $(\bX,\bY,\bZ,\bF)$, test statistic $T(\bx, \by, \bz, \mathbf{f})$, total number of re-samples $B$\;
  \For{$b = 1, 2, \dots, B$}{
  Sample $\bF^b$ uniformly from $\Pi_J$ the set of all permutations on $\{1,\dots,J\}$;
 }
 \textbf{Output:} $p$-value $:= \frac{1}{B+1} \left[1 + \sum_{b=1}^{B} \mathbbm{1}\{T(\bX,\bY,\bZ, \bF^b) \geq T(\bX,\bY,\bZ, \bF)\}\right]$
 \caption{CRT: Fatigue Effect} \label{algo:CRT_fatigue}
\end{algorithm}

\section{Details of the Simulation Setup}
\label{appendix:simulation}

We provide here a detailed description of our simulation setup. For simplicity, we focus on the setting in which each respondent only has one evaluation, i.e., $J = 1$. Appendix~\ref{appendix:multiple_tasks_sim} (see Figure~\ref{fig:RE_sim}) presents additional simulations that have multiple evaluations per respondent based on respondent random effects. In the setting where each respondent evaluates only one task, we treat each response as an independent observation and drop subscript $j$.  We also assume that every factor $(\bX,\bZ)$ is uniformly and independently randomized, implying that all treatment combinations are equally likely. As before, $\bX$ represents the main factors of interest while $\bZ$ denotes the other factors. For simplicity, we assume that all factors ($\bX, \bZ$) are binary with their success probabilities equal to 0.5, and we have one factor of interest ($q = 1$).

\subsection{The Basic Setup}
\label{appendix:sim_basic_setup}
To clearly separate main and interaction effects, we use the sum-to-zero constraint by coding each binary factor as $(-0.5, 0.5)$.  Our data generating process uses the following logistic regression model under the forced-choice design,
\begin{align*}
& \Pr(Y_{i} = 1 \mid X_i, \bZ_i) \ = \ \text{logit}^{-1}\left[\beta_{X} (X_{i}^{L} - X_{i}^{R}) + \beta_{Z}^\top (\bZ_{i}^{L} - \bZ_{i}^{R}) \right.\\
& \quad \quad  + 2 \gamma^\top \{(X_{i}^{L}\bZ_{i}^{L}) - (X_{i}^R\bZ_{i}^{R})\} +\left. 2 \delta^\top \{(X_{i}^{L} \bZ_{i}^{R}) - (X_{i}^{R}  \bZ_{i}^{L})\} + 2 \tilde{\gamma}^\top \{(\bZ_{i}^{L} \times \bZ_{i}^{L}) - (\bZ_{i}^R \times \bZ_{i}^{R})\} \right],
\end{align*}
where $\beta_{X}$ and $\beta_{Z}$ represent the coefficient vectors for the main effects of $\bX$ and $\bZ$, respectively, and $\gamma$ and $\delta$ denote the coefficient column vectors for the within-profile and between-profile interactions between $\bX$ and $\bZ$, respectively. For simplicity, we omit between-profile interactions among $\bZ$ and consider only within-profile interactions among $\bZ$ with effect sizes $\tilde{\gamma}$. To facilitate interpretation, each interaction coefficient is multiplied by $2$ because our encoding of interaction effects, e.g., $\{(X_{i}^{L}\bZ_{i}^{L}) - (X_{i}^R\bZ_{i}^{R})\}$, results in three possible values $(-0.5, 0, 0.5)$. 

This data generating process also implies the absence of profile order effects. Lastly, we note that the logistic regression has a latent variable, $Y_i'$, representation such that $Y_i = 1 \text{ if } Y_i' > 0$ and 0 otherwise, where we let $\epsilon_i$ follow a standard logistic distribution and
\begin{align*}
Y_i' &= \beta_{X} (X_{i}^{L} - X_{i}^{R}) + \beta_{Z}^\top (\bZ_{i}^{L} - \bZ_{i}^{R}) \\
& + 2 \gamma^\top \{(X_{i}^{L} \bZ_{i}^{L}) - (X_{i}^R \bZ_{i}^{R})\} + 2 \delta^\top \{(X_{i}^{L} \bZ_{i}^{R}) - (X_{i}^{R} \bZ_{i}^{L})\} + 2 \tilde{\gamma}^\top \{(\bZ_{i}^{L} \times \bZ_{i}^{L}) - (\bZ_{i}^R \times \bZ_{i}^{R})\}+ \epsilon_i.
\end{align*}
We consider settings in which $\bX$ has one main effect of $\beta_{X} = 0.1$, which is fixed for all simulations. In addition, $\bZ$ consists of ten factors with four non-zero main effects with a magnitude of $0.1$ with alternating signs, which is fixed for all simulations, i.e.,  $\beta_{Z} = (0.1, -0.1 ,0.1, -0.1, 0, \dots, 0)$. Lastly, we fix $\tilde{\gamma}^\top$ to have fifteen non-zero entries of 0.05, where the non-zero interactions are randomly chosen from all possible within-profile interactions among $\bZ$. The remaining entries are all zero. 

The sample size is fixed to $n = 3,000$ throughout the simulations. For each simulation, we generate within-profile and between-profile interactions between $\bX$ and $\bZ$ by randomly selecting the specified number of interactions from all possible interactions. The total number of non-zero interaction effects between $\bX$ and $\bZ$ varies from 0 to 18. The number of non-zero within-profile interactions between $\bX$ and $\bZ$ is kept identical to that of non-zero between-profile interactions between $\bX$ and $\bZ$. We make all non-zero within-profile interactions positive and all non-zero between-profile interactions negative while fixing all non-zero interaction effects between $\bX$ and $\bZ$ to be equal in magnitude. We explore additional simulations in Appendix~\ref{appendix:sim_heterogenous} where there are heterogeneous interaction effects. We set $B = 200$ for all simulations presented in this paper. 

We measure the total variance with respect to the latent representation of the logistic regression. Consequently, we define the variance explained by the interaction effects between $\bX$ and $\bZ$ and all other ``remaining'' effects (main effects of $\bX$, $\bZ$, and interactions among $\bZ$) as,
\begin{align*}
    \sigma^2_{\text{Interaction}} & := \mathbb{V}(2 \gamma^\top \{(X_{i}^{L} \bZ_{i}^{L}) - (X_{i}^R \bZ_{i}^{R})\} + 2 \delta^\top \{(X_{i}^{L} \bZ_{i}^{R}) - (X_{i}^{R} \bZ_{i}^{L})\}) \\
    \sigma^2_{\text{Remaining}} & := \mathbb{V}(\beta_{X} (X_{i}^{L} - X_{i}^{R}) + \beta_{Z}^\top (\bZ_{i}^{L} - \bZ_{i}^{R}) + 2 \tilde{\gamma}^\top \{(\bZ_{i}^{L} \times \bZ_{i}^{L}) - (\bZ_{i}^R \times \bZ_{i}^{R})\}),
\end{align*}
where $\mathbb{V}(\cdot)$ denotes the variance of the respective random variable. The total variance is defined as $\sigma^2_{\text{Interaction}} + \sigma^2_{\text{Remaining}}$. Since $\beta_X, \beta_{Z}, \tilde{\gamma}$ are fixed for all simulations, $\sigma^2_{\text{Remaining}}$ is fixed with a value of:
$$\sigma^2_{\text{Remaining}} = 9\times 0.1^2 \times 2\mathbb{V}(X_{i}^{L}) + 15 \times 4 \times 0.05^2 2 \mathbb{V}(X_{i}^{L} X_{i}^{R}) = 9\times 0.1^2 \times \frac{1}{2} + 15\times 4 \times 0.05^2 \times \frac{1}{8} = 0.06375,$$
where the first equality holds because all random variables $(X_i^L, X_i^R, \bZ_i^L, \bZ_i^R)$ are independent, centered at zero, and identically distributed (element-wise identically distributed for the multivariate $\bZ_i^L$ and $\bZ_i^R$). 

Furthermore, given a signal size $I$ and the number of interactions $n_{I}$ ($n_I$ denotes the total number of within-profile and between-profile interactions), we have that:
$$\sigma^2_{\text{Interaction}}  = 8I^2\mathbb{V}(X_i^{L}X_i^{R})n_{I} = \frac{I^2 n_{I}}{2}.$$
Finally, since the left plot of Figure~\ref{fig:interaction} contains six interactions ($n_I = 6$), the $x$-axis is computed by $\frac{3 I^2}{3 I^2 + 0.06375}$, where $I$ takes the following values $(0, 0.025$, $0.05, 0.075, 0.1, 0.125)$.
\vspace{0.1cm}

To calculate the statistical power of each test, we compute a $p$-value for each of 1,000 Monte Carlo data sets.  For the CRT, we use the HierNet test statistic given in Equation~\eqref{eq:hiernet} and impose the constraints in Equation~\eqref{eq:constraints}, whereas we use the $t$-test based on the estimated regression coefficient of $\bX$ for the AMCE-based test. The AMCE-based analysis assumes no profile-order effect. Therefore, as suggested by \cite{AMCE}, we run the linear regression by stacking all the left and right profiles, resulting in $2n$ rows. More formally, we have the new response $\bY^{\text{AMCE}} = [\bY; \mathbf{1} - \bY]$ regressed on $\bX^{\text{AMCE}} = [\bX^L; \bX^R]$, where $\bX^L = [X_1^L; X_2^L; \dots; X_n^L]$ and $\bX^R$ is defined similarly (note we have dropped the $j$ subscript since $J = 1$). Finally, the standard errors are clustered by respondents as suggested by \citeauthor{AMCE}. Since $J = 1$ for our simulation, we cluster the standard errors on each evaluation task, i.e., each unique cluster consists of the left and right profile for each task. We then compute the power as the proportion of $p$-values that are less than $\alpha = 0.05$.  

\subsection{Heterogeneous Interaction Size}
\label{appendix:sim_heterogenous}

As stated in Section~\ref{section:simulations}, we fix all interaction sizes to be equal for every simulation. Here, we examine if the simulation results presented in Figure~\ref{fig:interaction} change if there are heterogeneous interaction effects. 

We compare the statistical power under two different data generating processes using the CRT with test statistic in Equation~\eqref{eq:hiernet}. We denote the original data generating process as the ``homogeneous'' scenario since all interaction sizes are equal under this scenario. We create an additional  ``heterogeneous'' scenario that contains two unique varying interaction effects - one that is strong, $I_s$, and one that is weak $I_w$. To facilitate a fair comparison between the ``heterogeneous'' and ``homogeneous'' scenario, we force the total variance explained by the interactions to be equal under both scenarios. We assign all strong interaction effects to the within-profile interaction and all weak interaction effects to the between-profile interaction. Suppose there are only two non-zero interactions between $\bX$ and $\bZ$. Since we impose $\sigma^2_{\text{Interaction}}$ to be equal under both the ``homogeneous'' and ``heterogeneous'' scenario, we have the following equation:
$$I_w^2 + I_s^2 = 2I^2,$$
where $I$ is the interaction size for the ``homogeneous'' scenario.

The possible values of strong and weak effects lie on the circle of radius $\sqrt{2}I$ centered at the origin. We pick $I_s(I) = \sqrt{\frac{4}{3}}I$ and $I_w(I) = \sqrt{\frac{2}{3}}I$, i.e., the point corresponding to the thirty degree angle of the circle. The variance explained by the interaction is equal for both the ``heterogeneous'' and ``homogeneous'' scenario when there are two non-zero interactions. To create a power curve for the ``heterogeneous'' scenario analogous to the left plot in Figure~\ref{fig:interaction}, we create three interactions of size $I_s(I)$ for the within-profile interaction and three interactions of size $I_w(I)$ for the between-profile interaction. The two scenarios still maintain equal variance explained by the interaction effects because all random variables are independent and centered at zero. For the analogue of the right plot of Figure~\ref{fig:interaction}, we similarly keep the relative proportion of $I_w(I)$ and $I_s(I)$ fixed and increase the number of non-zero interactions to match the ``homogeneous'' scenario. For example, if there are twelve non-zero interaction effects, then there are six within-profile interactions of size $I_s(I)$ and six between-profile interactions of size $I_w(I)$ for the ``heterogeneous'' scenario. 

\begin{figure}[t]
\begin{center}
\includegraphics[width=\textwidth]{"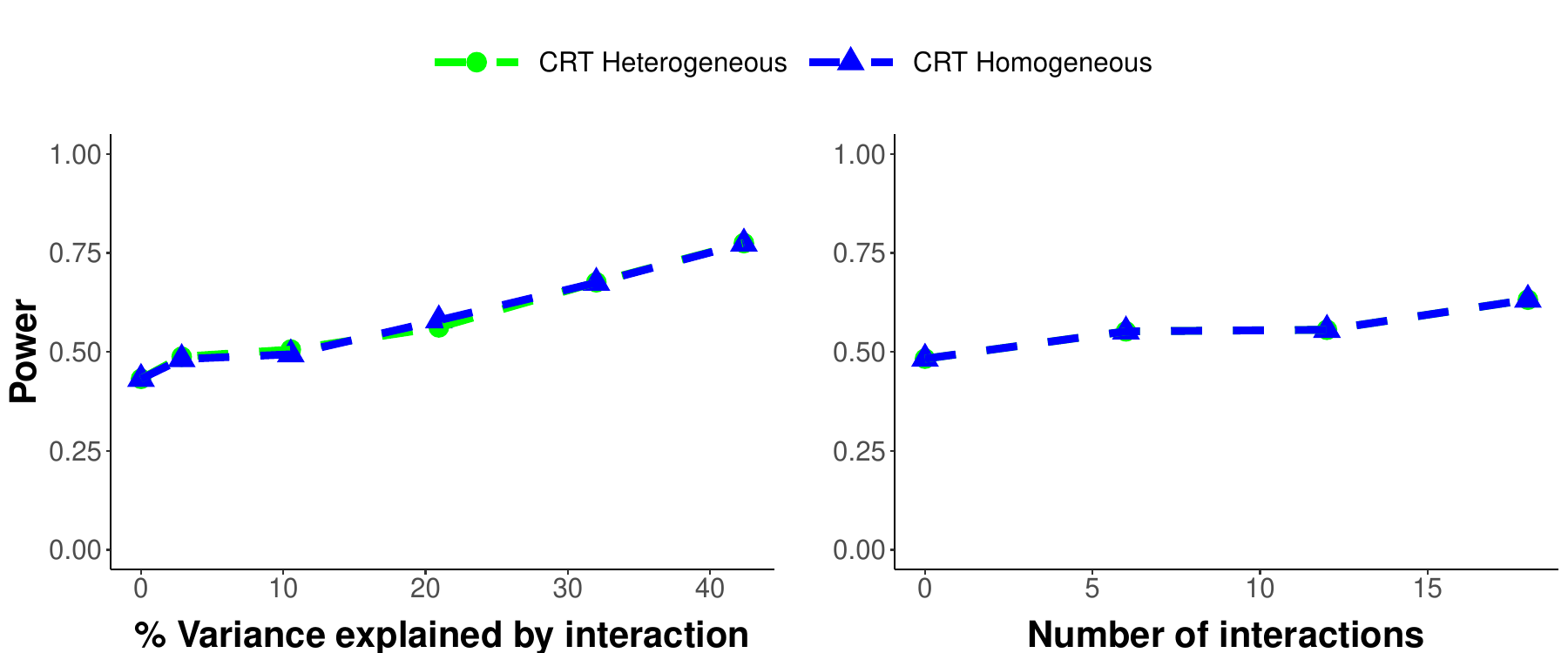"}
\caption{The figure shows the power of the ``heterogeneous'' (light green circles) and ``homogeneous'' (blue triangles) scenario in the same simulation setting as in Figure~\ref{fig:interaction}.}
\label{fig:hetero}
\end{center}
\end{figure}

Figure~\ref{fig:hetero} shows that the power of the ``heterogeneous'' and ``homogeneous'' scenario is indistinguishable under the same simulation setting in Figure~\ref{fig:interaction}. This shows that we lose no generality by considering only the simple ``homogeneous'' scenario in the main simulations in Figure~\ref{fig:interaction}.

\subsection{Additional AMCE Simulations}
\label{appendix:AMCE_sim}

\begin{figure}[t]
\begin{center}
\includegraphics[width=\textwidth]{"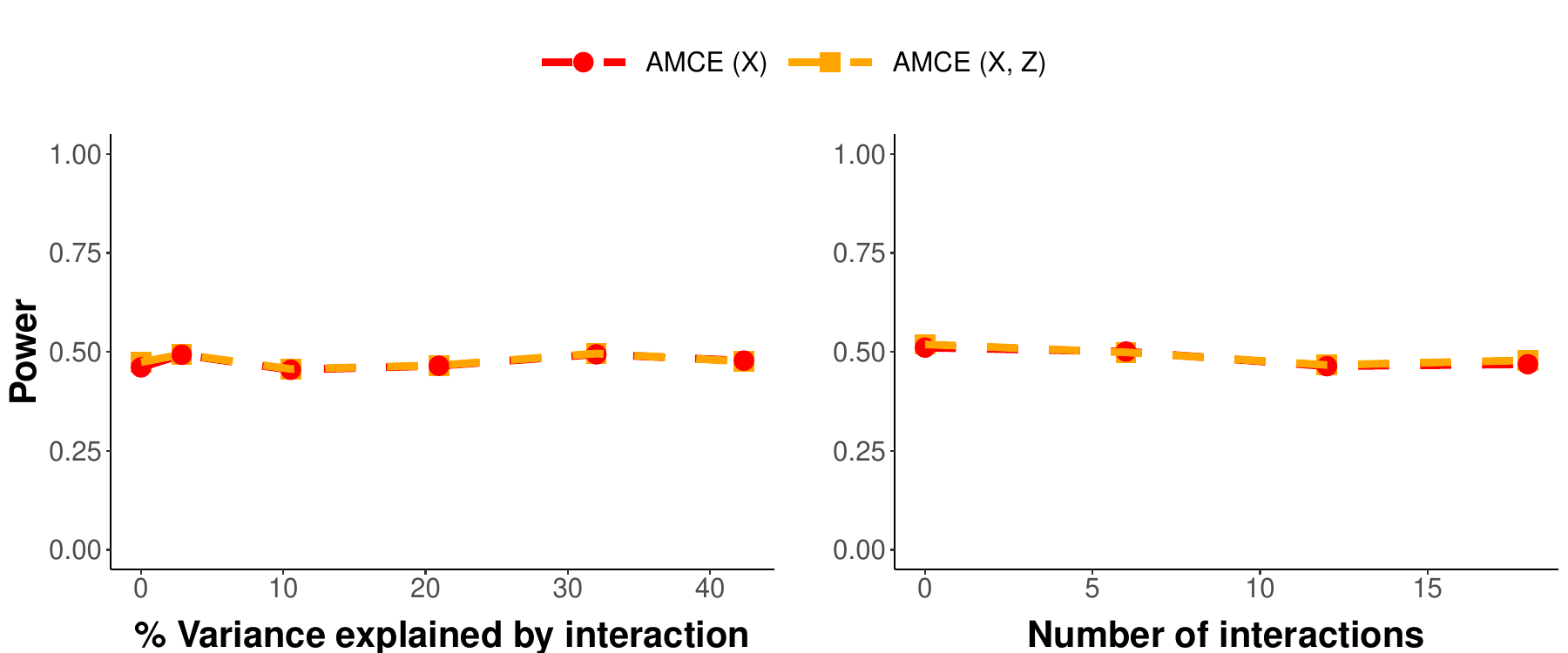"}
\caption{The figure shows the power of the ``long AMCE'' that uses all $(\bX,\bZ)$ in the linear regression fit (orange squares) and the original AMCE that uses only $\bX$ in the linear regression fit (red circles) in the same simulation setting as in Figure~\ref{fig:interaction}.}
\label{fig:AMCE_long}
\end{center}
\end{figure}

The AMCE computed in Figure~\ref{fig:interaction} was based on a linear regression of $\bY$ on $\bX$, without $\bZ$ included among the predictors. Although this is valid and sufficient to compute the AMCE of $\bX$, practitioners often compute the AMCE of all factors ($\bX, \bZ$) simultaneously with a single linear regression of $\bY$ on all $(\bX, \bZ)$. We compute the power of the ``long AMCE'' that is based on the $t$-test for the estimated regression coefficient of $\bX$ obtained by regressing the response $\bY$ on all $(\bX, \bZ)$. Figure~\ref{fig:AMCE_long} shows that the power of the ``long AMCE'' (orange squares) is indistinguishable from that of the original AMCE presented in Figure~\ref{fig:interaction} (red circles). 

\subsection{Simulations with Multiple Tasks per Respondent}
\label{appendix:multiple_tasks_sim}

\begin{figure}[t]
\begin{center}
\includegraphics[width=\textwidth]{"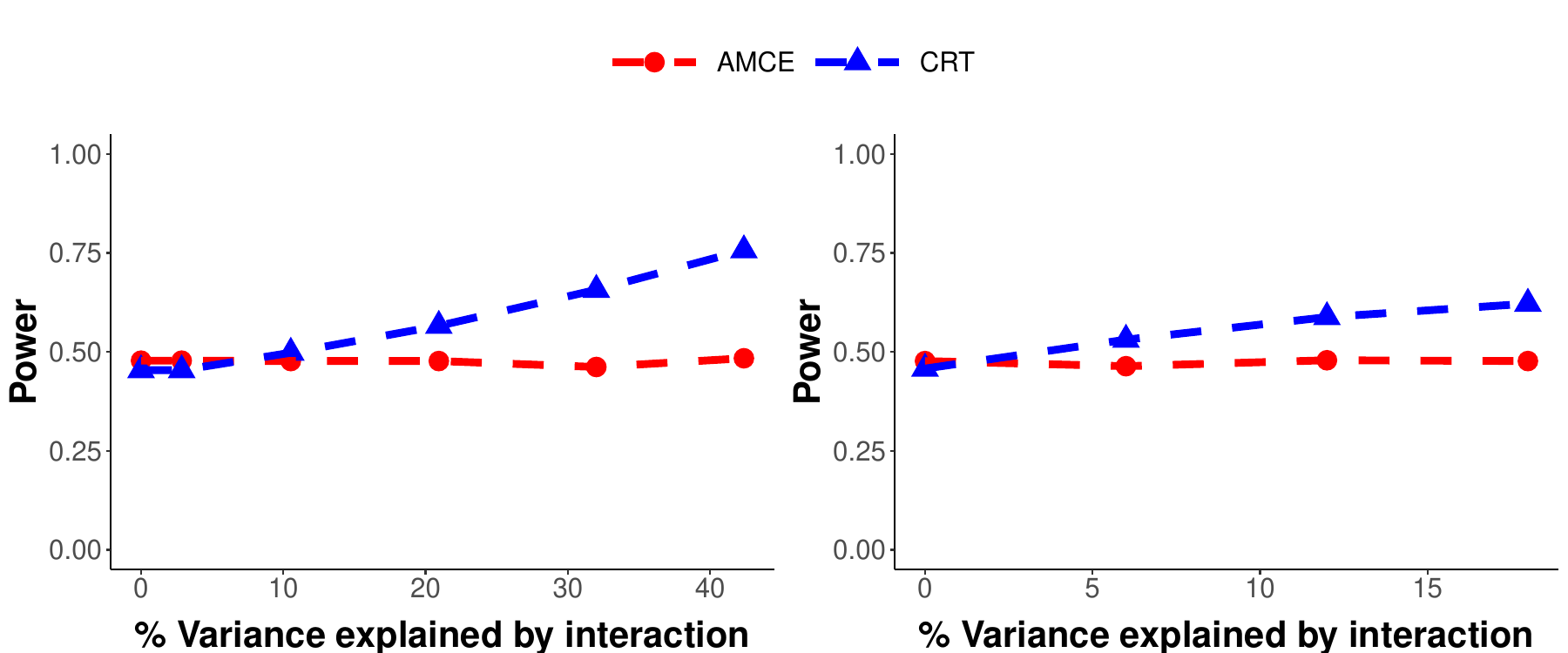"}
\caption{The figure shows the power of the AMCE (red circles) and the CRT (blue triangles) using the HierNet test statistic in Equation~\eqref{eq:hiernet}. We modify the simulation setting in Figure~\ref{fig:interaction} by having each respondent evaluate $J = 5$ tasks with a total of $nJ = 3,000$ responses. Otherwise, the simulation setup remains identical to that in Figure~\ref{fig:interaction}. Each respondent has a random effect of $\sigma_{RE}^2 = 0.1$.}
\label{fig:RE_sim}
\end{center}
\end{figure}

Appendix~\ref{appendix:sim_basic_setup} presents the simulation results where each respondent evaluates only one task ($J = 1$).  Here, we consider a simulation setup where each respondent evaluates $J = 5$ tasks while still fixing the total sample size $nJ = 3,000$. We keep the same simulation setup as the one described in Appendix~\ref{appendix:sim_basic_setup} except that we allow each respondent to have a random effect $U_j \sim N(0, \sigma_{RE}^2)$. More formally, our new data generating process is, 
\begin{align*}
& \Pr(Y_{ij} = 1 \mid X_{ij}, \bZ_{ij}) \ = \ \text{logit}^{-1}\left[\beta_{X} (X_{ij}^{L} - X_{ij}^{R}) + \beta_{Z}^\top (\bZ_{ij}^{L} - \bZ_{ij}^{R}) \right.\\
& \quad  + 2 \gamma^\top \{(X_{ij}^{L}\bZ_{ij}^{L}) - (X_{ij}^R \bZ_{ij}^{R})\} +\left. 2 \delta^\top \{(X_{ij}^{L} \bZ_{ij}^{R}) - (X_{ij}^{R}  \bZ_{ij}^{L})\} + 2 \tilde{\gamma}^\top \{(\bZ_{ij}^{L} \times \bZ_{ij}^{L}) - (\bZ_{ij}^R \times \bZ_{ij}^{R})\} + U_j \right],
\end{align*}
where $U_j$ is the random effect for each respondent $j$. We keep all simulation parameters the same as that in Figure~\ref{fig:interaction} except we use the above data generating process with $J = 5$ evaluation tasks and random effects with $\sigma_{RE}^2 = 0.1$ to produce Figure~\ref{fig:RE_sim}. 

Although the CRT HierNet test statistic does not change with the addition of multiple respondent evaluations, the AMCE estimate must properly account for the respondent effect. As suggested by \cite{AMCE}, we use the robust clustered standard errors clustered on respondents for the linear regression of $\bY$ on $\bX$ and use the $t$-test based on the estimated regression coefficient of $\bX$ to produce the power curve (red). Figure~\ref{fig:RE_sim} shows the results are similar to those shown in Figure~\ref{fig:interaction}, suggesting that our results are not sensitive to the number of evaluations per respondent. 

\section{Leveraging Presidential Data in the Gender Application}
\label{appendix:presidential_data}

We detail here how we leverage the Presidential candidate data to find the strongest interaction, as mentioned in Section~\ref{subsection:gender_results}. Because we are only interested in whether the additional interactions are significant, we run the Lasso logistic regression with main effects of $(\bX,\bZ,\bV)$ and one interaction between $\bX$ (``gender'') and one variable in $(\bZ,\bV)$, producing one CRT $p$-value for each variable in $(\bZ,\bV)$. For example, when including an interaction between ``gender'' and ``profession'', we include both within-profile and between-profile interactions between all levels of ``gender'' and ``profession''. Consequently, the test statistic for testing the interaction between $\bX$ (``gender'') and factor $\ell$ of $\bZ$ is:
\begin{equation}
T_{\text{Presidential}}^{\text{Candidate Factor, } \ell} = \sum_{k=1}^{K} \sum_{k^\prime = 1}^{K_{\ell} } (\hat\gamma_{1 \ell kk^\prime} - \bar\gamma_{1 \ell k^\prime})^{2} + \sum_{k=1}^{K } \sum_{k^\prime = 1}^{K_{\ell} } (\hat\delta_{1 \ell kk^\prime} - \bar\delta_{1 \ell k^\prime})^{2}.
\label{eq:lasso_profile}
\end{equation}
The test statistic for testing the interaction between $\bX$ and factor $m$ of the respondent characteristic $\bV$ is:
\begin{equation}
T_{\text{Presidential}}^{\text{Respondent Characteristic, } m} = \sum_{k=1}^{K } \sum_{w = 1}^{L_m } (\hat\xi_{1 m k w} - \bar\xi_{1 m w})^{2},
\label{eq:lasso_respondent}
\end{equation}
where all coefficient estimates refer to the same corresponding estimates in Equation~\eqref{eq:modified_hiernet} and $K = 2$ refers to levels \textit{male} and \textit{female}. Lastly, when running the CRT we similarly enforce the constraints in Equation~\eqref{eq:constraints} (along with the constraints on the respondent characteristics) by fitting the Lasso logistic regression on the appended $D^{c}$ data matrix.

Table~\ref{tab:lasso_presidential} shows the resulting $p$-value for each variable in $(\bZ,\bV)$.  Many of the variables have $p$-values lower than 0.1.  The variables such as ``position on immigrants'', ``position on abortion'', ``position on government deficit'', and ``position on national security'', are all related to the disparate views Democratic and Republican candidates may have, in line with ``party affiliation'' being the most significant. Even among the respondent characteristics, the respondent's political ideology, a measure of how conservative or liberal the respondent is, is the most  significant variable. We conduct a robustness analysis by repeating the analysis of Section~\ref{subsection:gender_results} but using the second most significant variable, the ``position on abortion'' factor, to interact and include as the additional main effect in HierNet, though we note that ``position on abortion'' is quite a bit less significant than ``party affiliation'' with more than four times as large a $p$-value. The resulting $p$-value is 0.078. Although this is not as significant as the main analysis, it still provides suggestive evidence that gender plays a role in voting for Congressional candidates. 

\begin{table}
\small
\begin{center}
\begin{tabular}{lc} 
\textbf{Variable} & \textbf{$p$-value}  \\ 
\hline
Age &  0.060 \\ 
Race & 0.31\\ 
Family &  0.22\\ 
Experience in public office & 0.30\\ 
Salient personal characteristic & 0.45\\ 
Party affiliation & 0.0049\\ 
Policy area of expertise & 0.25\\ 
Position on national security & 0.067\\ 
Position on immigrants & 0.067 \\ 
Position on abortion & 0.022 \\ 
Position on government deficit & 0.032 \\ 
Favorability rating among public &  0.63 \\ 
\hline
Respondent gender & 1.00\\ 
Respondent education & 1.00\\ 
Respondent age & 1.00\\ 
Respondent class & 0.41 \\ 
Respondent region & 1.00 \\ 
Respondent race & 1.00\\ 
Respondent partisanship & 0.11 \\ 
Respondent thought on Hillary Clinton & 1.00 \\ 
Respondent interest in politics & 0.43\\ 
Respondent political ideology & 0.042\\ 
\hline
\end{tabular}
\caption{Resulting $p$-values using the CRT Lasso logistic regression with main effects of $(\bX,\bZ,\bV)$ and an additional interaction between $\bX$ and one variable in $(\bZ,\bV)$ from the Presidential candidate data. The test statistic captures the interaction terms with each variable in $(\bZ, \bV)$ as shown in Equation~\eqref{eq:lasso_profile} and Equation~\eqref{eq:lasso_respondent}, respectively. All respondent characteristic variables are labelled with ``respondent''.}
\label{tab:lasso_presidential}
\end{center}
\end{table}

\section{Computational Details}
\label{appendix:dICRT}

The HierNet test statistic introduced in Equation~\eqref{eq:hiernet} is powerful but can be computationally expensive because the CRT requires a total of $B + 1$ cross-validated HierNet fits. To address this problem, we speed up HierNet in three ways. First, we reduce the default convergence tolerance for the optimization algorithm in the HierNet package from $10^{-6}$ to $10^{-3}$. Second, following the ``distillation'' idea introduced in \citep{dCRT}, we cross-validate the sparsity parameter lambda only through a HierNet fit of $\bY$ on $\bZ$ without involving any $\bX$. Because $(\bY, \bZ)$ remains constant for all $B + 1$ fits, we only need one cross-validation fit. Lastly, we initialize the starting parameters in the optimization algorithm with one HierNet fit that is uniformly and randomly chosen from the $B + 1$ HierNet fits. Since we uniformly choose one out of $B + 1$ HierNet fits as the initialization, this procedure still satisfies the exchangeability needed for the CRT's validity. Because many of the parameters estimated from the $B + 1$ different HierNet fits will likely be similar to each other, the initialization likely saves computation time. 

\begin{figure}[t]
\begin{center}
\includegraphics[width=\textwidth]{"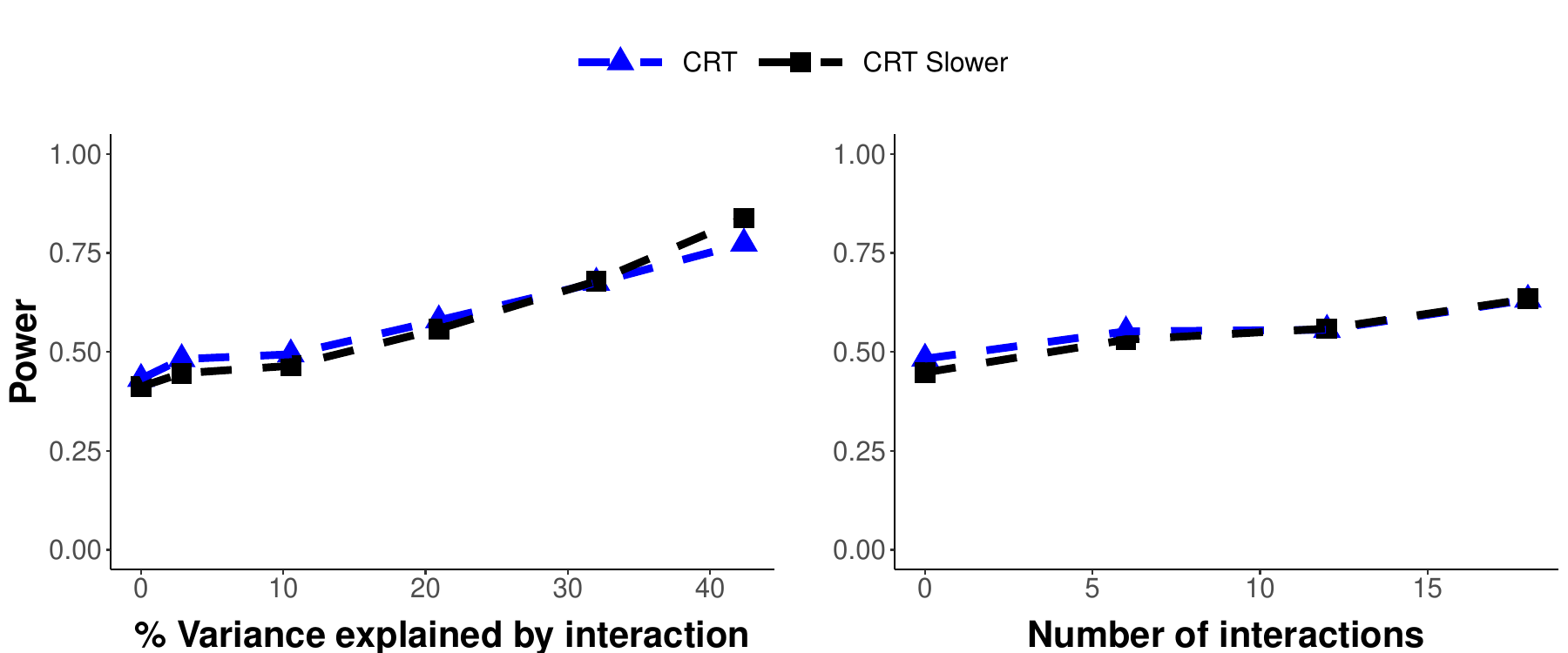"}
\caption{This figure represents the power of the original faster HierNet test statistic (blue triangles) with the three computational speedups and the slower HierNet test statistic (black squares) without the computational speedups in the same simulation setting as in Figure~\ref{fig:interaction}.}
\label{fig:dI_sim}
\end{center}
\end{figure}

Although the above procedure significantly reduces computational complexity, practitioners may worry if there is a significant loss of power from this simplification. Consequently, we plot in Figure~\ref{fig:dI_sim} the original HierNet power curve shown in Figure~\ref{fig:interaction} that leverages the aforementioned three speed-ups (in blue) and the computationally slower HierNet power curve without the three speed-ups (in black). Figure~\ref{fig:dI_sim} shows that the computational modifications have no significant impact on power. 

\section{Inflated $p$-values for Logistic Regression}
\label{appendix:logistic_reg}
\begin{figure}[t]
\begin{center}
\includegraphics[width=\textwidth]{"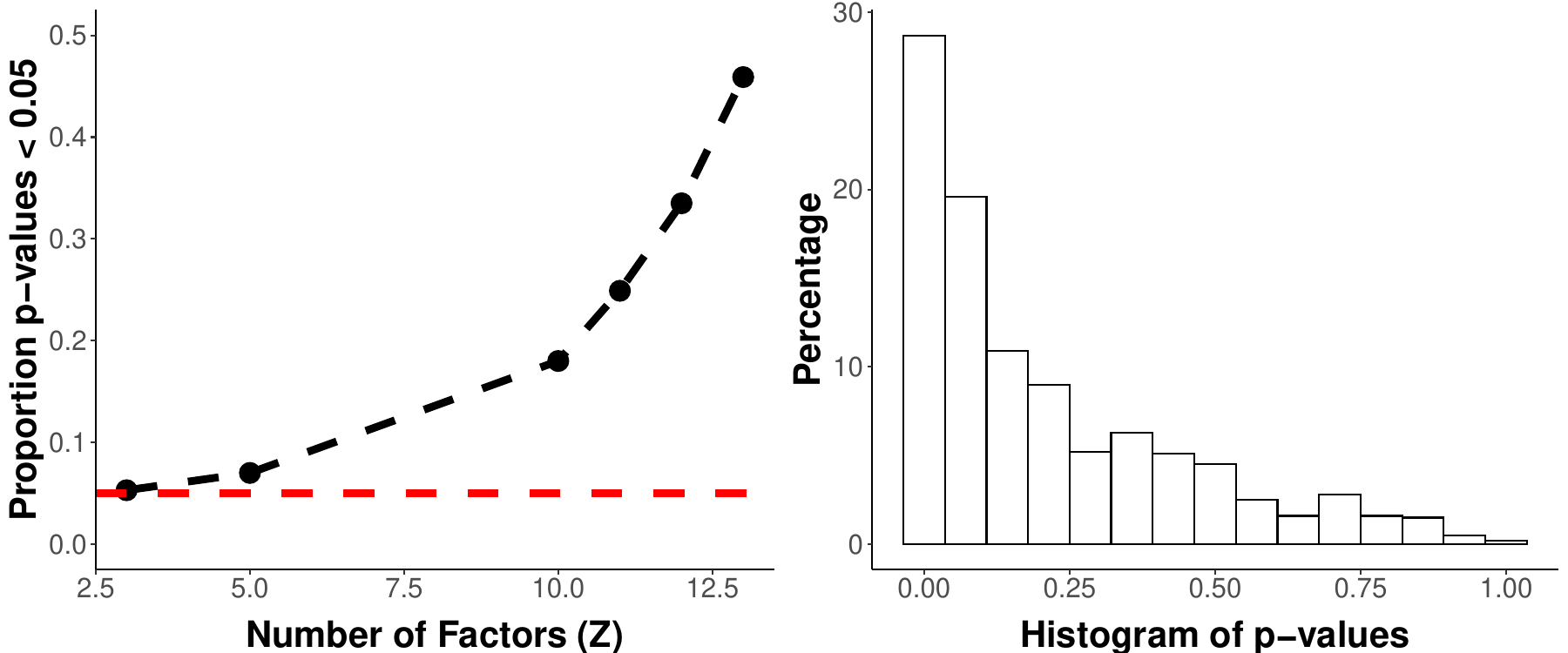"}
\caption{Inflated $p$-values from logistic regression. The left figure shows the proportion of $p$-values, obtained through a $F$-test from a logistic regression, less than $\alpha = 0.05$ when the number of other factors in $\bZ$ is $(3, 5, 10, 11, 12, 13)$ and $H_0$ is true. The red dotted line at $\alpha = 0.05$ represents the expected proportion of $p$-values less than $0.05$ if the $p$-values are valid. The right figure shows the histogram of 1,000 Monte-Carlo $p$-values when the number of other factors of $\bZ$ is 12. The sample size is $n = 5,000$ and each factor has four factor levels. All Monte Carlo standard errors are below $0.016$.}
\label{fig:appendix_logistic}
\end{center}
\end{figure}

Although the AMCE is popular in conjoint analysis, especially among political scientists, there also exist model-based approaches. Logistic regression remains a popular model-based approach to conjoint analysis \citep{mcfa:73, marketgeneral,marketreg}.  We explore in this section how this modeling approach can lead to invalid inference in conjoint analysis. When testing $H_0$, researchers may want to account for not only the main effects of $(\bX,\bZ)$ but also all two-way interactions, as done similarly in HierNet, to reduce model misspecification. Under this scenario, we show through simulations in Figure~\ref{fig:appendix_logistic} that even reasonable sample sizes and dimensions of $(\bX,\bZ)$ can lead to invalid $p$-values, i.e., the type 1 error is greater than the desired $\alpha \in [0, 1]$.

We use a similar simulation setting as the one in Appendix~\ref{appendix:sim_basic_setup} but simplify it further. Since we are interested in showing how the $p$-values obtained from a logistic regression may be invalid in general, we do not have ``left'' or ``right'' profiles but only one profile, leading to the following data generating process, 
\begin{align*}
& \Pr(Y_{i} = 1 \mid X_i, \bZ_i) \ = \ \text{logit}^{-1}\left[\beta_{X} X_i + \beta_{Z}^\top \bZ_i + \gamma^\top (X_{i}\bZ_{i}) +  \tilde{\gamma}^\top (\bZ_{i} \times \bZ_{i})  \right].
\end{align*}
All factors have four levels, and we similarly assume one factor of interest $q = 1$ while varying the number of other factors. Since we are interested in the behavior of the $p$-values under the null $H_0$, we force all effects of $X$ on the response to be zero, i.e., $\beta_X = \gamma = 0$. For simplicity we also make all effects of $\bZ$ zero, i.e., $\beta_Z = \tilde{\gamma} = 0$ and fix the sample size to $n = 5,000$. To reflect the researcher's desire to reduce model misspecification by accounting for all two-way interactions as HierNet does, we fit a logistic regression of $\bY$ on all main effects and two-way interactions of $(\bX,\bZ)$. We then obtain a $p$-value for testing $H_0$ by an $F$-test that tests $\beta_X = \gamma = 0$. We obtain 1,000 Monte-Carlo $p$-values and plot the proportion of $p$-values less than $\alpha = 0.05$ in the left plot of Figure~\ref{fig:appendix_logistic}. We also vary the number of factors of $\bZ$, which is shown in the $x$-axis of the left plot. On the right plot of Figure~\ref{fig:appendix_logistic}, we plot the histogram of the 1,000 $p$-values obtained when the number of factors of $\bZ$ is 12.

Under the null hypothesis, we expect any valid $p$-value to have type 1 error control, i.e., $P(p\text{-value} \allowbreak \leq \alpha) \leq \alpha$ for all $\alpha \in [0,1]$. The left plot of Figure~\ref{fig:appendix_logistic} shows that only five other factors of $\bZ$ is enough to cause the proportion of $p$-values less than $\alpha = 0.05$ to be noticeably inflated at 7\%. The inflation becomes particularly apparent when there are twelve other factors of $\bZ$, which causes the proportion of $p$-values less than $0.05$ to be as high as 34\%. The histogram on the right plot of Figure~\ref{fig:appendix_logistic} visually shows how the $p$-values are clearly far from the expected uniform distribution and have an undesirable peak at zero, resulting in poor type 1 error control. This phenomenon is studied in \citep{phasetransition} and arises because the $p$-values' validity in a logistic regression depends on a low-dimensional asymptotic result. We note that a conjoint analysis has typically more than ten factors, where each factor usually has more than three levels. Therefore, Figure~\ref{fig:appendix_logistic} shows the potential dangers of using a model-based approach like the logistic regression to flexibly capture all interactions.

\end{document}